\documentclass[reqno,a4paper]{amsart}
\usepackage{amssymb,stmaryrd}
\usepackage{amsfonts}
\usepackage{amstext}
\usepackage{color}
\usepackage{graphicx}
\usepackage{subcaption}
\usepackage{hyperref}
\usepackage{amsmath}

\newcommand{\extd}{\mathrm{d}}

\newcommand{\del}{\partial}

\newcommand{\R}{\mathbb{R}}

\newcommand{\CH}{\mathcal{H}}

\newcommand{\CD}{\mathcal{D}}

\newcommand{\eps}{\epsilon}

\renewcommand{\imath}{{\mathfrak{i}}}
\renewcommand{\div}{{\mathrm{div}}}

\begin{document}

\title{Geodesic flows on a black-hole background}

\author[Kaushlendra Kumar and Shahn Majid]{\large Kaushlendra Kumar and Shahn Majid \\ \ \\ 
  School of Mathematical Sciences\\ 
  Queen Mary University of London\\ 
  Mile End Road, London E1 4NS, UK}

\thanks{
{\it Authors to whom correspondence should be addressed:}  kaushlendra.kumar@qmul.ac.uk and s.majid@qmul.ac.uk}

\thanks{{\it Funding:} KK was supported by a DFG project grant 515782239 and SM by a Leverhulme Trust project grant RPG-2024-177}
\date{June 2025 Ver 1.2}



\date{\today}

\begingroup
\let\MakeUppercase\relax 
\maketitle
\endgroup

\author{Kaushlendra Kumar and Shahn Majid}

\begin{abstract} A recent notion of geodesic flows which comes out of noncommutative geometry but which is also novel in the classical case is studied in detail for a Schwarzschild spacetime. In this framework, the geodesic velocity field is an independent concept which then defines the flow of a density $\rho$ on spacetime or possibly that of an amplitude wave function $\psi$ with $\rho=|\psi|^2$. The proper time flow parameter $s$ is generated collectively by the flow of matter. We show carefully how the $\rho$ evolution can be justified as modelling a large number of geodesics interpolated as a local density. Using Kruskal-Szekeres coordinates, we show that there are no issues crossing the horizon. A novel feature is that whereas two colliding Gaussian bumps in density $\rho$ merge into a single bump, two colliding wave function $\psi$ bumps of opposite phase merge into a dipole with a different density $|\psi|^2$ profile, providing a potential test of our wave-function hypothesis. We also revisit the Klein-Gordon flow or pseudo-quantum mechanics around a black-hole and find that previously found black-hole atom states and modes generated at the horizon when an area of disturbance approaches it are also present inside the black-hole in a reflected fashion. We argue that the behaviour of the horizon modes across the horizon as well as discretization of the atomic spectrum depend on quantum gravity corrections at the horizon. 
\end{abstract}

\section{Introduction}

In noncommutative geometry, which is conjectured to be a better approximation than continuum geometry at the Planck scale\cite{Ma:pla,DFR,MaRue,Hoo}, one need not have any points, in which case we cannot have any geodesics either. A solution to this problem has been developed over several papers \cite{Beg:geo,BegMa:geo,LiuMa, BegMa:cur,BegMa:gra} and consists classically of thinking of a dust of particles with density $\rho$ which then evolves as each dust particle moves along its own geodesic. One can then let $\rho$ be an evolving positive element of a $*$-algebra $A$, or even go a step further in the direction of quantum theory and let $\rho=\psi^*\psi$ where $\psi\in \CH$, a Hilbert space completion of $A$, plays the role of a wave function or amplitude.

 In this paper, we explore this formalism in detail for a {\em classical} pseudo-Riemannian spacetime manifold $M$, specifically a black-hole background. The formalism has elements in common with (but is significantly different from) the standard methods of relativistic fluid mechanics\cite{Olt}  as well as the theory of optimal transport\cite{LotVel} and the theory of geodesic flows on the normal bundle in Riemannian geometry. The latter was famously applied to recover the Euler equation in fluid mechanics by  Arnold\cite{Arn66} and is also key to modern notions of ergodicity. The intuitive picture in our case is to imagine that the tangent vectors of all the dust particles provide a vector field, which we call the {\em geodesic velocity vector field} $X$. This turns out, however, to obeys its own equation that does not even mention $\rho$, namely the {\em geodesic velocity equation},
\begin{equation} \label{velocityflow} 
\dot X + \nabla_X X=0, 
\end{equation}
where the dot denotes $\extd\over\extd s$. Next, $X$ indeed relates to the rate of change of $\rho$,  according to the {\em density flow equation},
\begin{equation}\label{densityflow}
\dot\rho + X(\rho) + \rho\, \div(X)=0
\end{equation}
but here again there is a surprise: this equation determines $\dot \rho$ from $X$ but not the other way around. Thus, the concept of a quantum geodesic, applied here in the classical case, rips apart the usual notion of a geodesic, where the position comes first and its motion determines the velocity, and glues these back in reverse order, where  $X$ is a field in its own right that then determines the flow of any density $\rho$. Our first goal in the present work is to understand the meaning of such geodesic velocity fields $X$, i.e., how one should choose them and think about them physically given that they are the more fundamental object. A second goal is to see what happens if we go further and write $\rho=|\psi|^2$ as in quantum mechanics, where $\psi$ evolves as
\begin{equation}\label{ampflow} 
\dot\psi+ X(\psi) + {1\over 2}\psi\, \div(X) =0.
\end{equation}
This {\em amplitude flow equation} implies the density flow equation and is what is suggested by the formalism of quantum geodesics when applied to a classical manifold. Note that if $\psi(0)$ is real then $\psi$ remains so under this evolution as all the terms of the PDE are real, but it could still be oscillatory. If we write $\psi=\sqrt{\rho}e^{\imath\theta}$ then it is easy to see that the remaining content of the amplitude flow equation is 
\begin{equation}\label{angflow} 
\dot \theta+X(\theta)=0
\end{equation}
for which $\theta(0)=0$ stays zero. This is, however, a special case as typical $\psi$ do not have such a polar decomposition in a smooth way. (For example, a real $\psi$ that crosses zero will have a jump discontinuity in $\theta$.) Note that all of these equations feature the {\em convective derivative} along $X$ as in fluid mechanics, defined as
\begin{equation}
  \frac{D}{D s}|_X f= \dot f+ X(f),\quad   \frac{D}{D s}|_X Y =  \frac{\extd Y}{\extd s} + \nabla_X Y
\end{equation}
for a function $f$ or another vector field $Y$. It is known \cite{BegMa:cur} that the length of $X$ and its divergence then obey 
\begin{equation}\label{eq:derNormXsq}
    \frac{D}{D s}|_X (|X|^2) = 0 ,\quad 
\frac{D}{D s}|_X (\mathrm{div}(X)) = -(\nabla_\mu X^\nu) (\nabla_\nu X^\mu) - X^\mu X^\nu R_{\mu\nu},  
\end{equation}
in local coordinates, where $R_{\mu\nu}$ is the Ricci tensor. The second of these fits the meaning of the Ricci tensor in general relativity (GR) as controlling how a body changes shape (or the stress it experiences) as it freely falls. Meanwhile,  the first of (\ref{eq:derNormXsq}) implies that if $|X|^2=\pm 1$ at $s=0$ then it remains so at later times, since $X(1)=0$. Hence we can limit attention to unit speed velocity fields, which is  a little closer (but not the same as) the notion of a geodesic flow on the unit normal bundle in differential geometry. We will, moreover, work in the Lorentzian case with signature $-+++$, so for timelike flows we actually set $|X|^2=-1$. We also add to our considerations the  flow of densities and amplitudes. 

We will be solving these equations in a black-hole background, with $\nabla$ the Levi-Civita connection and the Ricci tensor zero other than at the singularity in the centre. The goal, building on \cite{LiuMa} in the Minkowski spacetime case, is to better understand the physical interpretation, given the counter-intuitive elements above about the role and meaning of $X$ in its own right. While we can easily imagine an initial density $\rho$ to evolve, for example, a Gaussian bump at $s=0$, what do we take for $X$? The mathematical  answer is that each geodesic starts with a point and a velocity chosen independently, so we can choose any $X$ at $s=0$, but we still need to chose it in a physical way. Our proposal for this is that for any vector field $X$, we consider the infinitesimal diffeomorphism generated by it. If $\rho$ is a function on $M$ then $\rho(s)(x)= \rho(x- s X(x))$ 
can be viewed as the corresponding change of any density, for small $s$, in which case $\dot\rho(s)(x)=-X(\rho(s))$, i.e. ${D\over D s}|_X(\rho)=0$ at $s=0$. Hence for this kind of motion, we deduce from the density flow equation (\ref{densityflow}), for nonzero densities, that 
\begin{equation}\label{divcon} \div(X(0))=0.\end{equation}
By the second of (\ref{eq:derNormXsq}) we know that this will not typically stay zero, but it provides a natural condition, along with the unit speed condition, for our starting point of the geodesic flow. This and suitable boundary conditions relevant to the matter content will, in our case (where the model is effectively 2-dimensional as we focus on the radial-time sector of the theory) then determine the initial $X(0)$.

We will follow this line, but a more physical answer could be that the initial matter distribution at $s=0$ is governed by the stress-energy tensor. For a perfect fluid, this contains density, pressure and a velocity field with $|u|^2=-1$ in the form
\[ T_{\mu\nu}= (\eps+p)u_\mu u_\nu + p g_{\mu\nu},\]
where it is usual to write this in terms of a number density $n$ (in the case of one particle type) and a chemical potential $\mu$ according to
\[ \eps+p=\mu n,\quad \extd\eps=\mu \extd n.\]
In this case, part of the conservation law $\nabla_\mu T^{\mu}{}_\nu=0$ is the equation $\div(n u)=0$. If we set $X=u$ and $n=\rho$ then we see that $\dot\rho=0$. One also has the negative acceleration obeying
\[ -\nabla_u u={(\extd p)^\#+ u\, u(p)\over \eps + p}\]
where $\#$ turns a 1-form into a vector field via the metric. The above identification would then say that $\dot u$ is determined by the expression on the right hand side, i.e. a nonlinear evolution equation for $u$. These observations do not necessarily contradict relativistic fluid mechanics, because evolution with respect to an external time $s$ is not considered there. They give a special case where $\rho$ does not evolve but $X$ does and both have a physical interpretation. Moreover, the derivation of $\div( n u)=0$ comes out of an equation of state and could potentially be relaxed.

Another issue is the `proper time parameter' $s$, which amounts to a kind of collective emergent time from the geodesic flow and is intrinsic to the flow. Geometers have no problem imagining a single particle at $s=0$ and letting it evolve with respect to its own proper time $s$, but doing this probabilistically for every point of the manifold amounts to a collective time which needs to be understood further. To explore these issues, our first result, in Section~\ref{sec:stat}, is to actually evolve a large number `dust' of geodesics sprinkled with a Gaussian density distribution and show that it indeed compares well with the quantum geodesic flow theory applied classically. Another unusual aspect is the use of wave functions $\psi$, as these live in $L^2(M)$ over the whole spacetime $M$. They evolve with respect to $s$ and hence should not be thought of as quantum mechanics in any usual sense (which would be relative to a foliation by time-slices).  In Section~\ref{sec:examples}, we look at two such Gaussian bumps colliding using the quantum geodesic flow formalism, at this $\psi$ level. We find that while two density bumps collide and merge into a single bump, in the $\psi$ case where the bumps have different phase (we take one bump positive and the other negative) the result of the collision is a dipole with a different split-bump density profile. Thus, the hypothesis from quantum geodesics\cite{Beg:geo, BegMa:geo, BegMa:cur}  that densities $\rho$ could be replaced by an underlying wave functions can in principle be tested, once there is a better handle on the physical interpretation of $s$ and of spacetime wave functions.    

Section~\ref{sec:kg} rounds off the paper with a look at a different quantum geodesic flow which is closer to actual quantum mechanics, namely  on the noncommutative algebra $\CD(M)$ of differential operators\cite{BegMa:geo, BegMa:qm,BegMa:flrw} as  a coordinate-invariant version of the Heisenberg algebra. The `Schr\"odinger picture' version of the flow also entails evolving  `wave functions' $\phi\in L^2(M)$ defined over spacetime (we use a different symbol to avoid confusion with the amplitude $\psi$ above) but now obeying the {\em Klein-Gordon flow} 
\begin{equation}\label{KGflow}
-\mathrm{i} \frac{\partial \phi}{\partial s} = \frac{\hbar}{2m} \square \phi,
\end{equation}
where \(\square\) is the Klein-Gordon operator. Now $s$ as proper time parameter is justified by operator-level `Heisenberg picture' geodesic flow equations
\begin{align}
m \frac{\mathrm{d} x^\mu}{\mathrm{d} s} &= g^{\mu \nu} p_\nu - \frac{\lambda}{2} \Gamma^\mu, \label{momenta} \\
m \frac{\mathrm{d} p_\mu}{\mathrm{d} s} &= \Gamma^\nu_{\mu \sigma} g^{\sigma \rho} \left( p_\nu p_\rho - \lambda \Gamma^\tau_{\nu \rho} p_\tau \right) + \frac{\lambda}{2} g^{\alpha \beta} \Gamma^\nu_{\beta \alpha, \mu} p_\nu, \label{momentaFlow}
\end{align}
in first order phase space form and in local coordinates where $\lambda=-\imath\hbar$ and $[x^\mu,p_\nu]=\imath\hbar\delta^\mu{}_\nu$ . Here  \(g^{\mu \nu}\) is the metric, \(\Gamma^\mu = \Gamma^\mu_{\alpha \beta} g^{\alpha \beta}\) are contracted Christoffel symbols, and $m$ is a mass parameter. Stationary states of (\ref{KGflow}) are solutions of the massive Klein-Gordon equation with mass determined by the eigenvalue, but the theory is not limited to such on-shell modes and in this respect carries some of the information one might expect from QFT on the background.  

Of interest is the case where $M$ has a static metric, in which case one can look at $\phi$ of a fixed energy with respect to the coordinate time evolution and a remaining function $\psi$ on spatial slices. Then (\ref{KGflow}) reduces to {\em pseudo-quantum mechanics}, which is similar to but not the same as ordinary quantum mechanics because there is no one observer whose time is $s$. A discovery in \cite{BegMa:qm} using Schwarzschild co-ordinates were certain `horizon modes' emanating from the horizon when an area of disturbance reaches it. We show that one has the same feature in Kruskal-Szekeres coordinates outside the event horizon, as well as atom-like stationary modes with the black-hole in the role of nucleus. This confirms that the theory is indeed generally covariant as claimed in \cite{BegMa:qm}, but this time we can go inside the black-hole where we find similar behaviour but reflected. Approaching the horizon from either side, the horizon modes and atom-like solutions oscillate infinitely fast (they have a fractal form) but we argue that quantum gravity corrections would cut off the highest frequencies and thereby allows modes and evolution of flows up to and across the horizon. Using the numerical resolution of the finite element PDE solver to give a first impression of what this could look like, it would appear that horizon modes penetrate in a thin skin to the other side of the horizon. As physical checks, the evolution of $\psi$ in both Sections~\ref{sec:examples} and Section~\ref{sec:kg} is unitary (i.e. preserving the norm) to within numerical accuracy, and the state entropy as the horizon modes are generated increases both for modes outside as in \cite{BegMa:qm} and inside the black-hole. 

We provide some concluding remarks in Section~\ref{sec:rem}. We note that noncommutative geodesic flows also appear in Connes `spetral geometry' in terms of an abstract Dirac operator \cite{Con95}. Numerical flows are done with Mathematica NDSolve using the Method of Lines. This has a discretisation parameter res:=MaxCellMeasure for the maximum cell size of the spatial geometry which we will vary in order to check robustness against changes of resolution. Restricting MaxStepSize for the evolution parameter $s$ did not improve plots whenever we tried it, so this was left to be chosen automatically. The code used to generate the plots is made available in \cite{KM:code}. 

\goodbreak

\section{Preliminaries on Kruskal-Szekeres coordinates and geodesics }\label{sec:pre}

We briefly recap Schwarzschild spacetime and associated conformal transformations adapted to the geometry near the black-hole horizon. The Schwarzschild metric for a non-rotating black-hole in Lorentzian signature \((-+++)\) and natural units ($G=c=1$) is,
\begin{equation}
\extd s^2 = -\left(1 - \frac{r_s}{r}\right) \extd t^2 + \left(1 - \frac{r_s}{r}\right)^{-1} \extd r^2 + r^2 \extd\Omega^2, \label{eq:schwarzschild_metric}
\end{equation}
where
\begin{equation}
r_s = 2M \quad \mathrm{and} \quad \extd\Omega^2 = \extd\theta^2 + \sin^2\theta \extd\phi^2. \label{eq:schwarzschild_params}
\end{equation}
Here $r_s$ is the Schwarzschild radius, and the coordinates are $-\infty < t < \infty$, $0 < r < \infty$, $0 \leq \theta \leq \pi$, $0 \leq \phi < 2\pi$. These coordinates are singular at the event horizon at $r = r_s$ and instead we use Kruskal-Szekeres coordinates, which extend the Schwarzschild metric analytically beyond the horizon and indeed into two other parallel regions. We first change to timelike coordinate \( T \) and spacelike coordinate \( X \) as follows (the angular coordinates are  untouched).

For the exterior regions \( (r > r_s) \):
\begin{equation}
T = \left( \frac{r}{r_s} - 1 \right)^{1/2} e^{r/(2r_s)} \sinh\left( \frac{t}{2r_s} \right), \quad
X = \left( \frac{r}{r_s} - 1 \right)^{1/2} e^{r/(2r_s)} \cosh\left( \frac{t}{2r_s} \right) .
\end{equation}

\begin{figure}
    \centering
    \includegraphics[width=0.6\textwidth]{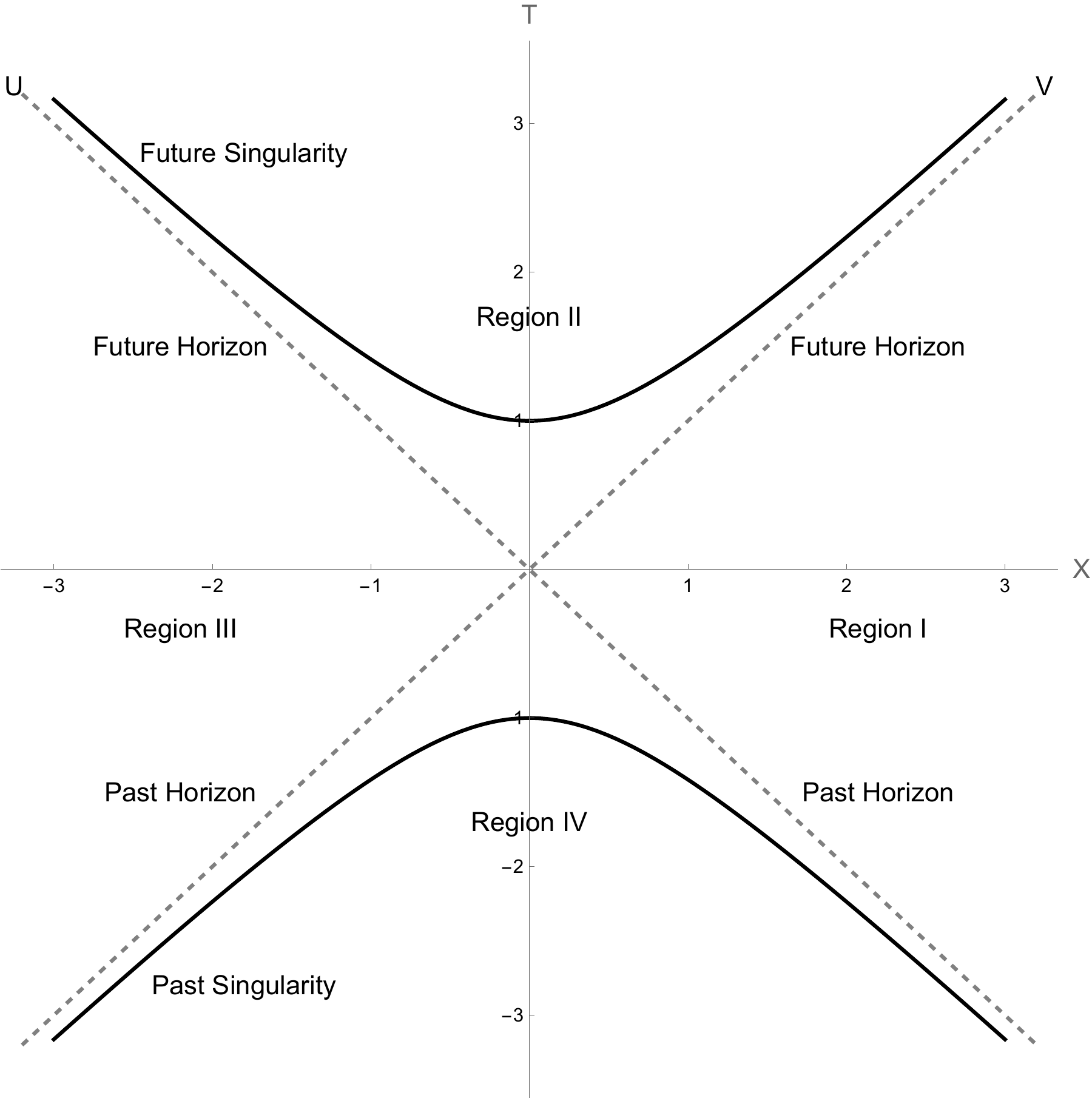}
    \caption{Kruskal-Szekeres diagram illustrating the four regions of spacetime: Region I (exterior), Region II (black-hole interior), Region III (parallel universe exterior), and Region IV (white hole interior). The event horizons are at the  dashed axes and the singularities are on the hyperbolas shown.}
    \label{fig:KShor}
\end{figure}

For the black-hole interior regions \( (0 < r < r_s) \):
\begin{equation}
T = \left( 1 - \frac{r}{r_s} \right)^{1/2} e^{r/(2r_s)} \cosh\left( \frac{t}{2r_s} \right), \quad
X = \left( 1 - \frac{r}{r_s} \right)^{1/2} e^{r/(2r_s)} \sinh\left( \frac{t}{2r_s} \right) .
\end{equation}
The relation \( T^2 - X^2 = \left( 1 - \frac{r}{r_s} \right) e^{r/r_s} \) implicitly defines \( r \) via the Lambert $W_0$ (principal branch) function:
\begin{equation}
r = r_s \left( 1 + W_0\left( \frac{X^2 - T^2}{e} \right) \right) . \label{rTXdef}
\end{equation}
As shown in Figure~\ref{fig:KShor}, the Kruskal diagram in the \( T \)-\( X \) plane has four quadrants separated by the event horizons now appearing as  \( T = \pm X \):
\begin{itemize}
    \item Region I (right quadrant): \( |T| < |X|, X > 0 \); the asymptotically flat exterior of our universe (\( r > r_s \)), where observers can escape to infinity.
    \item Region II (top quadrant): \( T > |X|, T > 0 \); the black-hole interior (\( 0 < r < r_s \)), from which nothing can escape, leading to the future singularity.
    \item Region III (left quadrant): \( |T| < |X|, X < 0 \); another asymptotically flat exterior (\( r > r_s \)), interpreted as a parallel universe.
    \item Region IV (bottom quadrant): \( T < |X|, T < 0 \); the white hole interior (\( 0 < r < r_s \)), a time-reversed black-hole from which matter emerges, connected to the past singularity.
\end{itemize}
The $r=0$ singularity is now spread  along the hyperbolas \( T^2 - X^2 = 1 \) . General  curves of constant \( r \) are further hyperbolas, while constant \( t \) are straight lines through the origin.  A standard Kruskal diagram showing these quadrants can be seen in Figure \ref{fig:KShor}. 

Finally, we move to null coordinates \( U = T - X, V = T + X \) where the $U,V$ axes are the event horizons. Using tortoise coordinates \( u = t - r^*, v = t + r^* \), where \( r^* = r + r_s \log \left| \frac{r}{r_s} - 1 \right| \), the explicit expressions for \( U \) and \( V \), adjusted for signs in each quadrant to ensure consistency, are:
\begin{itemize}
    \item In region I: \( U = -\exp\left( -\frac{u}{2r_s} \right), V = \exp\left( \frac{v}{2r_s} \right) \)
    \item In region II: \( U = \exp\left( -\frac{u}{2r_s} \right), V = \exp\left( \frac{v}{2r_s} \right) \)
    \item In region III: \( U = \exp\left( -\frac{u}{2r_s} \right), V = -\exp\left( \frac{v}{2r_s} \right) \)
    \item In region IV: \( U = -\exp\left( -\frac{u}{2r_s} \right), V = -\exp\left( \frac{v}{2r_s} \right) \)
\end{itemize}
These ensure \( U < 0, V > 0 \) in region I; \( U > 0, V > 0 \) in region II; etc. The relations are:
\begin{equation}
UV = -\left( \frac{r}{r_s} - 1 \right) e^{r/r_s}, \quad \frac{U}{V} = -\exp\left( -\frac{t}{r_s} \right) .\label{UVrelations}
\end{equation}
The metric in these coordinates comes out to be,
\begin{equation}
\extd s^2 = \frac{4r_s^3}{r} e^{-r/r_s} (-\extd U \extd V) + r^2 (\extd\theta^2 + \sin^2 \theta \extd\phi^2) .\label{UVmetric}
\end{equation}
It should be noted that this metric is off-diagonal in the $U-V$ sector with $g_{UV}=g_{VU}=-\frac{2r_s^3}{r} e^{-r/r_s}$. The  non-zero Christoffel symbols for the Levi-Civita connection in these coordinates are
\begin{equation}    \label{ChristoffelUV}
\begin{aligned}
     \Gamma^{U}_{UU} &= \frac{r_s^2}{r^2} ( 1 + \frac{r}{r_s}) V e^{-\frac{r}{r_s}},\quad \Gamma^{U}_{\theta\theta} = -\frac{U}{2} \frac{r}{r_s}, \quad \Gamma^{U}_{\phi\phi} = -\frac{U}{2} \frac{r}{r_s} \sin^2\theta, \\
    \Gamma^{V}_{VV} &= \frac{r_s^2}{r^2}( 1 + \frac{r}{r_s} ) U e^{-\frac{r}{r_s}}, \quad \Gamma^{V}_{\theta\theta} = -\frac{V}{2} \frac{r}{r_s},\quad \Gamma^{V}_{\phi\phi} = -\frac{V}{2} \frac{r}{r_s} \sin^2 \theta, \\
    \Gamma^{\theta}_{U\theta} &= \Gamma^{\theta}_{\theta U} = -V\frac{r_s^2}{r^2}e^{-\frac{r}{r_s}}, \quad \Gamma^{\theta}_{V\theta} = \Gamma^{\theta}_{\theta V} = -U\frac{r_s^2}{r^2}e^{-\frac{r}{r_s}}, \quad \Gamma^{\theta}_{\phi\phi} = -\sin\theta\cos\theta, \\
    \Gamma^{\phi}_{U\phi} &= \Gamma^{\phi}_{\phi U} = -V\frac{r_s^2}{r^2}e^{-\frac{r}{r_s}}, \quad \Gamma^{\phi}_{V\phi} = \Gamma^{\phi}_{\phi V} = -U\frac{r_s^2}{r^2}e^{-\frac{r}{r_s}}, \quad \Gamma^{\phi}_{\theta\phi} = \Gamma^{\phi}_{\phi\theta} = \cot\theta.
\end{aligned}
\end{equation}

The geodesic equations for a curve $x^\alpha(s)$ parameterized by an affine parameter $s$ (proper time for timelike geodesics) takes the form,
\begin{equation}
\frac{\extd^2 x^\alpha}{\extd s^2} + \Gamma_{\mu\nu}^\alpha \frac{\extd x^\mu}{\extd s } \frac{\extd x^\nu}{\extd s} = 0, \label{eq:geodesic_eq}
\end{equation}
where $\Gamma_{\mu\nu}^\alpha$ are the Christoffel symbols derived from the metric. The corresponding geodesic equations for radial motion in the Kruskal-Szekeres coordinates (\( U, V \)) can be written in terms of the four-velocity components as $\dot\theta=\dot\phi=0$ remaining zero and
\begin{equation}
\ddot U +  \frac{r_s^2}{r^2} ( 1 + \frac{r}{r_s}) e^{-\frac{r}{r_s}} V \dot U^2= 0, \label{eq:geodesicU}
\end{equation}
\begin{equation}
\ddot V +  \frac{r_s^2}{r^2} ( 1 + \frac{r}{r_s}) e^{-\frac{r}{r_s}} U \dot V^2 = 0. \label{eq:geodesicV}
\end{equation}
Here, dot denotes $\extd\over\extd s$. The radial coordinate $r=r(U,V)$ \eqref{rTXdef} is given in terms of the `product-log' or Lambert function $W$ as
\begin{equation}
    r(U,V) = r_s(1 + W(-UV/e)). \label{rUV}
\end{equation}

Null geodesics (light rays) follow 45-degree straight lines in (\(T,X\))-plane and correspond to $\dot U=0$ (\( U = \text{constant} \)) or $\dot V=0$ (\( V = \text{constant} \)), with the coordinates allowing a smooth crossing of the horizon at \( r = r_s \). 

\section{Comparison of density of actual geodesics with geodesic flow}
\label{sec:stat}

In this section, we establish the credibility of the quantum geodesic formalism applied classically by showing that it tracks the statistical interpolation of the motion of a bunch of actual geodesic dust particles. 

\subsection{Statistical model of many geodesics as density $\rho_{\rm stat}$ and velocity $X_{\rm stat}$}
\label{sec:rhoXstat}

\begin{figure}
    \centering
        \includegraphics[scale=0.8]{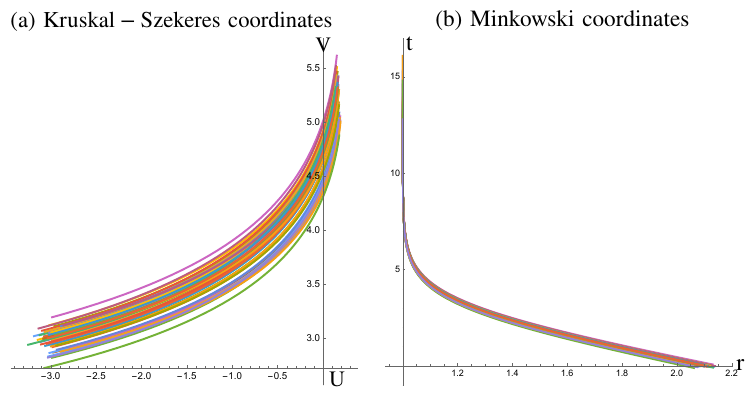}
    \caption{Parametric plots of geodesics for a bunch of $50$ nearby particles in (a) Kruskal-Szekeres and (b) Schwarzschild coordinates.} 
    \label{fig:geodBunch}
\end{figure}

We first set up a bunch of geodesics falling into the horizon $U=0$ in the Kruskal-Szekeres coordinates. We will set $r_s=1$ from now onwards for convenience. Next, we chose the initial ($s=0$) positions for $N$ number of massive dust particles, $(U_i(0),V_i(0))$ with $i=1,\ldots,N$ labeling each geodesic, by randomly sampling them from a normal distribution
\begin{equation}
    \rho(0)(U,V) := \frac{1}{2\pi\sigma^2\sqrt{-g}} \exp^{-\frac{(U+3)^2+(V-3)^2}{2\sigma^2}}\label{gauss0}
\end{equation}
 centered around $(-3,3)$ with standard deviation $\sigma=0.1$, and normalised so that $\int \sqrt{-g}\extd U\extd V \rho(0)=1$. We used  the measure on $(U,V)$-subspace
 \begin{equation}\label{measureUV}
    \sqrt{-g} = r(U,V)^2 |g_{UV}| = 2r(U,V)e^{-r(U,V)}
\end{equation}
from \eqref{UVmetric} in order to match the geometric picture, but this is equivalent to a Gaussian on the $U-V$ plane.  We also have to choose the initial velocities $(\dot U_i(0),\dot V_i(0))$ all with unit speed so as to admit interpolation as a smooth vector field. The simplest way to ensure this is to choose an actual initial unit norm vector field $X=(X^U,X^V)$, satisfying (being timelike)
\begin{equation}\label{unitSpeed}
    |X|^2:=2g_{UV}X^UX^V = -1,
\end{equation}
at all $(U,V)$ and set 
\begin{equation}
 \dot U_i(0)= X^U(U_i(0),V_i(0)), \quad  \dot V_i(0)= X^V(U_i(0),V_i(0)). \label{initveli}
\end{equation}
We choose a particular one with $X^V>0$ and $X^U>0$ (see Figure~\ref{fig:Xstat}) so that each geodesic is future directed and initially moves towards the horizon. With these initial conditions the particles in the bunch evolve with respect to $s$ following (\ref{eq:geodesicU}-\ref{eq:geodesicV}). We solve this numerically (using Mathematica) with results for $N=50$ geodesics shown as parametric plots in Figure \ref{fig:geodBunch} for both Kruskal-Szekeres and Schwarzschild coordinates, where the latter is singular at the horizon as expected.

\begin{figure}
    \centering
        \includegraphics[width=\linewidth]{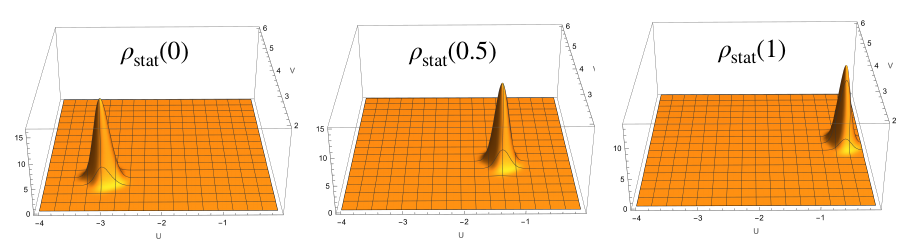}
    \caption{Probability density $\rho_{\rm stat}$ of the geodesic bunch of $N=50$ particles at $s=0,0.5,1$ showing motion to the right.}
    \label{fig:psiEvol}
\end{figure}

Next, having set up our bunch of geodesics each evolving with proper time $s$, we use Gaussian interpolation to extract a statistical density profile
\begin{equation}
    \rho_{\rm stat}(s)(U,V) = \frac{1}{2\pi\sigma^2N\sqrt{-g}}\sum\limits_{i=1}^N \exp({-\frac{(U-U_i(s))^2+(V-V_i(s))^2}{\sigma^2}}),
\end{equation}
where we use the same standard deviation $\sigma$ as when randomly choosing the initial locations in the bunch. Here the particle number $N$ in the denominator represents averaging and we have again inserted the geometric measure to ensure the normalisation $\int \sqrt{-g}\extd U\extd V \rho_{\rm stat}(s)=1$. This is such that  $\rho_{\rm stat}(s)$ can be interpreted as a probability density at each $s$. In Figure~\ref{fig:psiEvol}, we plot $\rho_{\rm stat}(s)$  for $s=0, 1/2, 1$, showing its expected motion to the right as the particles in the bunch evolve.

\begin{figure}[b]
        \centering
        \includegraphics[scale=.8]{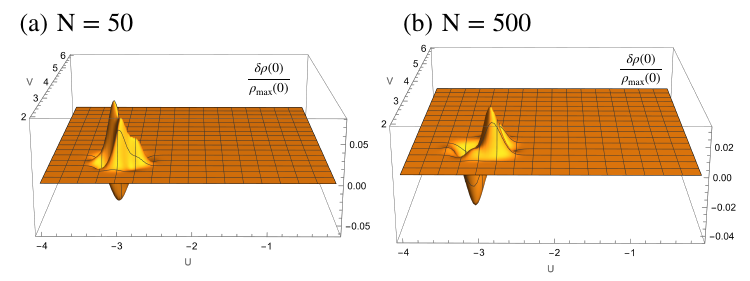}
    \caption{Difference $\delta\rho$ between the initial statistical $\rho_{\rm stat}(0)$ and the analytic $\rho(0)$ probability density as a fraction of the maximum of statistical $\rho_{\rm stat}(0)$ for (a) $N=50$ and (b) $N=500$ bunches.}
    \label{fig:rhodiff}
\end{figure}

\begin{figure}
\centering
\includegraphics[width=\linewidth]{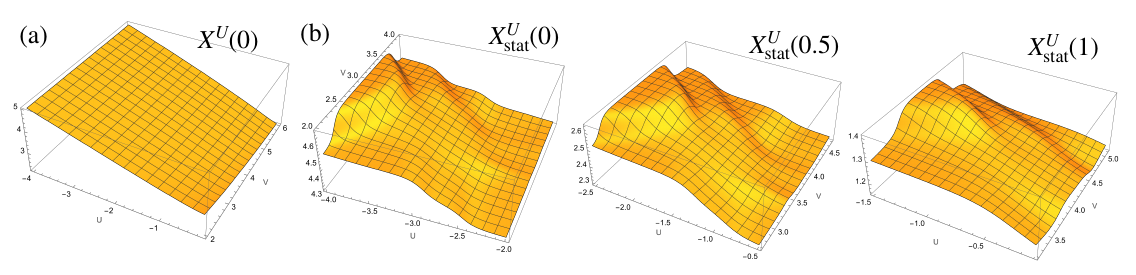}
\caption{Plots of (a) $X^U(0)$ used in the analytic construction, and (b) statistical $X^U_{stat}(s)$ for $N=500$. The plots for $X^V(0)$ and $X^V_{stat}(s)$ are similar.}
\label{fig:Xstat}
\end{figure}

To check the integrity of our methods, Figure \ref{fig:rhodiff} show how $\rho_{\rm stat}(0)$ compares with the actual analytic Gaussian $\rho(0)$ used to generate the geodesic start points, plotting their difference as a fraction of the maximum value $\rho(0)(-3,3)$. The center $(-3,3)$ here reflect the normal distribution with which the geodesic starting points were chosen. Also, the standard deviation is now $\sqrt{2}$ times that of the original normal distribution due to the Gaussian averaging of the sample. We see a good fit, that improves as the particle number increases to $N=500$, confirming that the differences here are a statistical fluctuation. These results are for the specific randomly-chosen bunches of  geodesics under study, but typical of others that we tried.

Finally, we make a similar statistical {\em but crude} interpolation of the tangents to each geodesic as a smooth vector field $X_{\rm stat}(s)$. Following the model of $\rho_{\rm stat}(s)$, we again use Gaussian interpolation, 
\begin{align}
    X^{U}_{\rm stat}(s)&= \frac{\sum\limits_{i=1}^N\rho_{\rm stat}(s)(U_i(s),V_i(s))X^{U}_i(s)e^{-\frac{(U-U_i(s))^2+(V-V_i(s))^2}{2\sigma^2}}}{\sum\limits_{i=1}^N\rho_{\rm stat}(s)(U_i(s),V_i(s))e^{-\frac{(U-U_i(s))^2+(V-V_i(s))^2}{2\sigma^2}}},
\end{align}
and similarly for $X^V_{\rm stat}$. This is crude because we are adding vectors at different points $(U_i,V_i)$ and to do this geometrically we should parallel transport to $(U,V)$. This is still some kind of approximation on the grounds that points near to $(U,V)$ contribute most. Although crude, this $X_{\rm stat}(s)$ serves to give a very rough idea of what kind of velocity field could be associated to a Gaussian bump of geodesics as they evolve. We see that $X_{\rm stat}(0)$ is not far from $X(0)$  in the vicinity of the statistical data points, but flattens out to its average value far away  (which does not have any geometric significance for the above reason). Figure \ref{fig:Xstat} also shows how $X^U_{\rm stat}(s)$ evolves at $s=0.5,1$  for later comparison. 

\subsection{Comparison of geodesic flow with $\rho_{\rm stat}(s)$ and $X_{\rm stat}(s)$}\label{sec:1bump}

\begin{figure}[b]
    \centering
        \includegraphics[scale=.8]{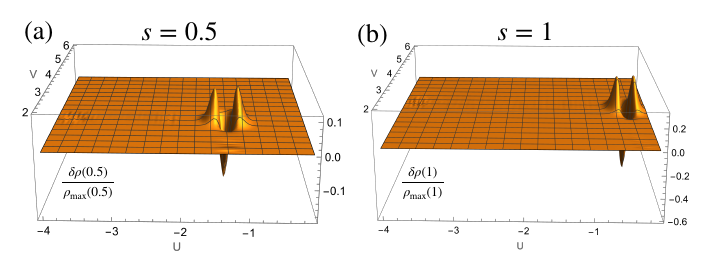}
    \caption{Difference at $s=0.5$ and $s=1$ between the geodesic flow model $\rho(s)$ and the statistical $\rho_{\rm stat}(s)$ as a fraction of maximum of $\rho_{\rm stat}$ for an $N=500$ bunch.}
    \label{fig:rhoEvol}
\end{figure}

In this section we compare the `observed' statistical flow $(\rho_{\rm stat}(s), X_{\rm stat}(s))$ with  $(\rho(s),X(s))$ that we get by starting with $\rho(0)$ and $X(0)$ as a smooth model of $\rho_{\rm stat}(0)$ and $X_{\rm stat}(0)$ respectively and evolving as a geodesic flow. For the latter evolution, the geodesic velocity equation for radial fields where $X^\theta(s)=X^\phi(s)=0$ and working in Kruskal-Szekeres coordinates, where the remaining components \eqref{velocityflow} for $X(s)=(X^U(s),X^V(s))$ depend only on $U,V$, is
\begin{equation}
\dot{X^U} + (\partial_U X^U)X^U + (\partial_V X^U) X^V + \frac{r(U,V)+1}{r(U,V)^2}e^{-r(U,V)}V(X^U)^2 = 0, \label{eq:geodFlowXu}
\end{equation}
\begin{equation}
\dot{ X^V} + (\partial_U X^V)X^U + (\partial_V X^V)X^V + \frac{r(U,V)+1}{r(U,V)^2}e^{-r(U,V)}U(X^V)^2 = 0,\label{eq:geodFlowXv}
\end{equation}
where the dot denotes differentiation with respect to $s$. The velocity equation for $X^\theta$ and $X^\phi$ states that these remain zero. If we focus on unit speed velocity fields \eqref{unitSpeed}, which, given the form of the metric, amounts to
\begin{equation}
    X^V = \frac{r(U,V)e^{r(U,V)}}{4X^U}\label{unitspeed}
\end{equation}
then (\ref{eq:geodFlowXv}) becomes redundant. Our broad strategy is therefore (i) to solve \eqref{eq:geodFlowXu} combined with \eqref{unitspeed} for some initial $X(0)$ to get the velocity flow $X(s)$ and (ii) use this to solve the corresponding density flow equation \eqref{densityflow}, which in Kruskal-Szekeres coordinates and in the radial case where there is no angular dependence, reads
\begin{equation}\label{densFlowUV}
    \dot\rho + \partial_U(\rho X^U) + \partial_V(\rho X^V) + \frac{r(U,V)-1}{r(U,V)^2}e^{-r(U,V)}\left(VX^U+UX^V\right)\rho = 0. 
\end{equation}

Specifically, we use $\rho(0)$ the Gaussian model for $\rho_{\rm stat}(0)$ and $X(0)$ the same smooth function that specified the initial velocities and which served as a model for $X_{\rm stat}(0)$ in Section~\ref{sec:rhoXstat}. We compute $X(s)$ numerically with boundary conditions $X(s)=X(0)$ at the boundary $U=-4$ (far away from horizon) and require $\rho(s)(-4,V)=\rho(s)(-0.1,V)=\rho(s)(U,0.1)=\rho(s)(U,6)=0$ at all four boundaries assuming that the area of disturbance never reaches them to any significant degree. We used  Mathematica with maximum cell measure res=$0.003$, sufficiently fine to be free from visible numerical noise. Before solving the density flow equation for $\rho(s)$ with respect to this $X(s)$ we do one thing purely to speed up the numerical solution at this point, namely we resample $X^U(s)$ on a deliberately coarser grid  with spacing $0.01$ in each of the $s,U,V$ variables (and determine $X^V(s)$ from this). Doing this reduces the computational load of using the interpolated functions $X^U(s),X^V(s)$ and their derivatives for the $\rho(s)$ PDE to something manageable, with no visible difference between the resampled $X(s)$ and the original one.

\begin{figure}
    \centering
      \includegraphics[width=\linewidth]{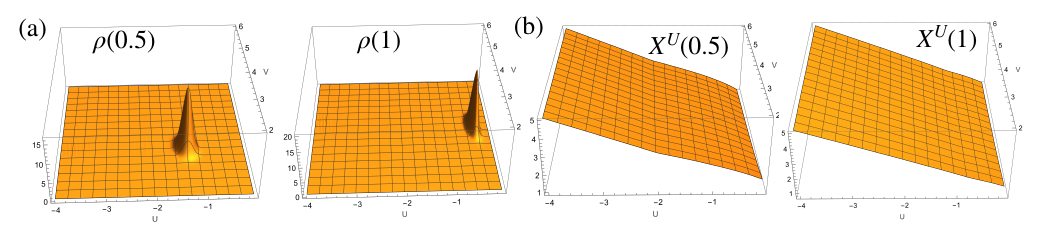}
    \caption{Geodesic flows for (a) density $\rho(s)$ and (b) velocity $X^U(s)$ at $s=0.5,1$}
    \label{fig:Xcompare}
\end{figure}

We find that the evolved $\rho(s)$ and $X(s)$ are a fair match to $\rho_{\rm stat}(s)$ and $X_{\rm stat}(s)$, i.e. the smooth flow follows the same trajectory as the interpolated bunch of geodesics. Here Figure~\ref{fig:rhoEvol} shows the relative difference $\delta\rho:=\rho_{\rm stat}-\rho$ at $s=0.5$ and $s=1$ (relative  to the max of $\rho_{\rm stat}(0)$ to set the scale) for $N=500$. One has the same picture and size of difference for $N=50$, suggesting that the difference is not of statistical origin. Rather, we see that it increases with $s$ presumably reflecting the changing shape of the peak compared to the statistical modelling which by construction produces Gaussian shape. Meanwhile, Figure \ref{fig:Xcompare} shows the evolved $X^U(s)$ at $s=0.5,1$. We already gave $\rho(0)$ as the initial Gaussian in (\ref{gauss0}) and  $X^U(0)$ in Figure~\ref{fig:Xstat}. Moreover, the integrity test of conserved density mass,
\begin{align}\label{integrity}
 \int \sqrt{-g}\extd U\extd V \rho(s)=1,
 \end{align}
 holds remarkably well during evolution with a numerical error of less than $0.002\%$ over the range to $s=1$.

\subsection{Choosing the initial velocity flow vector field}

Any unit speed time oriented vector field $X=X(0)$ can be used to prescribe the initial velocities of the geodesics and of the geodesic velocity equation. But how should we choose this physically? An approach which we use is to note\cite{LiuMa} that any vector field $X$ is an infinitesimal diffeomorphism of the manifold which evolves each point, and hence evolves all density functions $\rho$ on it in the opposite direction. As explained in the introduction, using the evolution of $\rho$ from (\ref{densityflow}) we obtain that $\div(X)=0$. What this initial condition corresponds to physically is a kind of uniform motion with no net flux of $X$ through any closed surface. We cannot necessarily impose this at all times $s$ but it provides a natural condition for the starting vector field $X(0)$. 

This divergence free condition combined with the unit-speed condition \eqref{unitspeed} (assuming $X^U,X^V>0$) gives
\begin{equation}\label{divfree}
4r^2(X^U)^2\del_U X^U + 4Ve^{-r}(r-1)(X^U)^3 = 2UrX^U + e^r r^3 \del_V X^U
\end{equation}
with $r=r(U,V)$ as before. Ordinarily, a PDE in two variables would require two boundary conditions, for example along the line segments $(-4, V)$ and $(U,6)$ as the boundary of the regions plotted above and where the corner (-4,6)  is far from the horizons. On the other hand, we can use this equation to express $\del_U X^U$ in terms of $X^U, \del_V X^U$. Hence one can specify a single Dirichlet boundary condition setting $X^U(U_0,V)$ along a line of constant $U=U_0$ and iteratively step through the values of $U_0$. Thus, 
$X^U(U_0+\eps,V)= X^U(U_0,V) + \eps \del_UX^U(U_0,V)+ O(\eps^2)$ and we replace the derivative by the values and $\del_V$ derivatives of $X^U(U_0,V)$ using (\ref{divfree}). This specifies $X^U(U_0+\eps,V)$ so as to numerically solve (\ref{divfree}) and we proceed iteratively. At the general step, will need to numerically compute $\del_V X^U$ along  the relevant line of constant $U$ (initially at $U=U_0$) and at the $V$-endpoints of that line we can take the 1-sided derivative from the interior, or equivalently, we can interpolate $X^U$ along the line to just beyond its endpoints to compute the derivative. Such an algorithm could be implemented, but a shortcut is to just take Neumann zero normal derivatives on the other line (here the line $(U,V_0)$) and then check directly that the solution indeed obeys $\div(X)=0$ away from from the boundary, just not exactly on the boundary. 

We focus on the case of  $X$ unit speed and covariantly constant along the main vertical line segment  $(U_0,V)$ at $U=U_0$ in the $U-V$ plane as the Dirichlet b.c.. For this we need $\nabla_{\del_V}X=0$, which using Christoffel symbol $\Gamma^V_{VV}|_{U=U_0}$ \eqref{ChristoffelUV} on $X^U(U_0,V), X^V(U_0,V)$ means
\begin{equation}
    \del_V X^U =0,\quad \del_V X^V + U_0 \frac{r(U_0,V)+1}{r(U_0,V)^2}e^{-r(U_0,V)}X^V = 0,
\end{equation}
that can be solved for the form of $r(U,V)$ \eqref{rUV} to get
\begin{equation}\label{XconstU0}
    X^U_{const\ U_0}(U_0,V)=\frac{c_1}{2},\quad X^V_{const\ U_0}(U_0,V) = \frac{1}{2 c_1} r(U_0,V)e^{r(U_0,V)},
\end{equation}
where we chose the second constant of integration so as to have unit speed. We do this along the line $(-4,V)$ with $c_1=10$ and the result was used in the previous section as shown in Figure~\ref{fig:Xstat}. We took the shortcut to impose Neumann zero on the `horizontal' line $(U,6)$ but note that $\del_VX^U_{const\ U_0}(U_0,V)$ with $U_0=-4$ is zero at the corner $(-4,6)$ which is consistent with Neumann along the line $(U,6)$ so there is no mismatch at the corner. This explains how we chose the field $X=X(0)$ used throughout Section~\ref{sec:stat}.

\begin{figure}
   \centering
       \includegraphics[width=\linewidth]{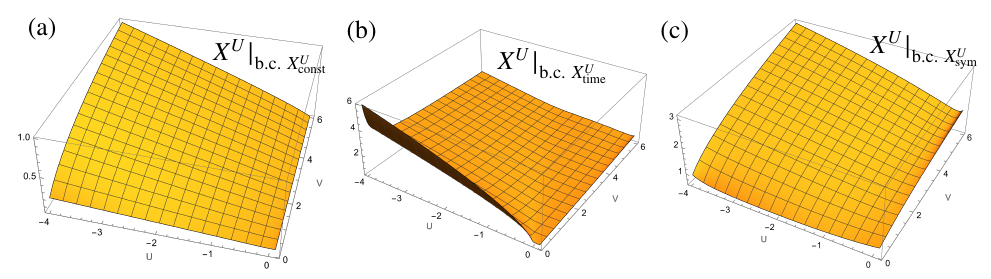}
    \caption{Other examples initial $X=X(0)$ obeying $\div(X)=0$ in the interior with the boundary values (a) $X_{const, V_0}$ on $(U,7)$, (b) $X_{time}$ and (c) $ X_{sym}$ on $(-4,V)$.\label{fig8}}
\end{figure}

If we wanted instead to impose Dirichlet b.c.s on the horizontal line $(U,V_0)$ of constant $V_0$, the covariance of $X$ along the vector field $\del_U$ translates to
\begin{equation}
    \del_U X^U + V_0\frac{r(U,V_0)+1}{r(U,V_0)^2}e^{-r(U,V_0)}X^U = 0,\quad \del_U X^V = 0,
\end{equation}
which can be solved  to get 
\begin{equation}\label{XconstV0}
    X_{const\ V_0}^U(U,V_0)= \frac{c_2}{2} r(U,V_0)e^{r(U,V_0)},\quad X^V_{const\ V_0}(U,V_0) ={1\over 2 c_2}.
\end{equation}
For example, imposing Dirichlet on $(U,6)$ and defaulting $(-4,V)$ to be Neumann we get the result shown in Figure~\ref{fig8}.  
Note that if we attempted to impose both Dirichlet conditions using 
\[ c_1= c_2 r(U_0,V_0)e^{r(U_0,V_0)}   \]
so that the values match at $(U_0,V_0)$, we would still have a mismatch in derivatives and no way to solve $\div(X)=0$ in the corner. Thus, $\del_U X^U(U_0,V)|_{V=V_0}$ computed from $X^U(U_0,V)$ and its $V$-derivatives by solving (\ref{divfree}) generally will not match $\del_U X^U(U,V_0)|_{U=U_0}$. Similarly, in general $\del_V X^U(U,V_0)|_{U=U_0}\ne \del_V X^U(U_0,V)|_{V=V_0}$. Indeed, imposing both Dirichlet b.c.s generically generates a kink artefact emanating from the corner and hence cannot be applied. 

Returning to our one Dirichlet b.c. along one edge (and Neumann on another) we can use other sources of the boundary field. The idea is that the boundary field is global data that then determines $X$ from $\div(X)=0$ in the bulk. We can fix it physically according to the flux or flux density of motion that we want to prescribe through the boundary edge. This will then steer the initial motion of all the particles in the bulk according to $X$ as $X(0)$. We can choose the boundary values as we please but one natural alternative to the covariantly constant case above is to note that $\del_t$ is a Killing vector as the metric is stationary. This is not unit speed but can be adjusted to  $X_{time} =(\sqrt{1-{1\over r}})^{-1}\del_t$ to be unit speed. On change of coordinates from \eqref{UVrelations}, this has components
\begin{equation}
    X^U_{time}=-{U\over 2\sqrt{1-{1\over r}}},\quad X^V_{time}={V\over 2\sqrt{1-{1\over r}}}.
\label{deltKilling}
\end{equation}
We then take the values of this as the boundary condition $(-4,V)$ and solve. The thinking is that  this is mostly far from the horizon and also far from the bump region of nontrivial density, i.e. where there is little matter. Hence at these boundaries we can expect $X(0)=X_{time}$ at the default timelike vector field in the absence of matter. The solution for $X(0)$ is shown in Figure~\ref{fig8}. Again, if we use $X_{time}$ to also impose a boundary condition at $(U,6)$ then we have a kink emanating from the corner, as an artefact. Meanwhile taking Neumann on $(U,6)$ is not quite right either but all this means is that $\div(X)\ne 0$ close to the border $(-4,V)$ as propagated from the corner $(-4,6)$ but this is not visible in $X_{time}$. 

Another natural unit speed vector field that may be of interest is to require $X^U=X^V$ which leads to 
\begin{equation}\label{Xsym} X^U_{sym}=X^V_{sym}={1\over 2} \sqrt{r}e^{r\over 2}\end{equation}
and again this could be used on the boundary $(-4,V)$ to determine the solution to $\div(X)=0$ with similar results as shown in Figure~\ref{fig8}, and again with a corner artefact if we impose both Dirichlet b.c.s. For all three cases of Figure~\ref{fig8} we have extended the $U$ domain to $U_{max}=0.1$  (which in principle can be extended further into Region II) to show that there is no particular problem at the horizon. 

Note that both $X_{time}$ and $X_{sym}$ could also be used directly for $X(0)$ in Region I as they are unit speed. Here, $X_{sym}$ has the added advantage of extending to other regions and also being non-singular at the curvature singularity at $r=0$.  For all fields, we  need $X(0)$ to be future-pointing at least in Region I,  which for unit speed field means 
\[ {X^U\over U}< {X^V\over V},\quad i.e.\quad (X^U)^2 - r^2{U\over 2 V \sqrt{-g}}>0 \]
where the second version assumes $X^U>0$. From this one can conclude that among unit speed fields in this region, positive time orientation holds if and only if $X^U,X^V>0$.

\section{Plots of analytic flows around a black-hole}
\label{sec:examples}

Having carefully verified our new geodesic flow methods compared to actual geodesics, we are now in position to explore some possible physics using them. We have seen that so long as the area of disturbance is localised, we can take $X(0)$ to be a single universal choice irrespective of the matter. In this case, its evolution $X(s)$ is also independent of the matter distribution and the density flow equation for $\rho(s)$ is then additive, i.e. a 2-bump geodesic flow is the sum of two 1-bump ones, and ditto for the amplitude flow for $\psi(s)$. The possibility that $\rho=|\psi|^2$ is a novel feature which we will be particularly interested in. The corresponding $U-V$ form of the amplitude flow in the radial case when there is no angular dependence and no angular components to $X(s)$ is 
\begin{equation}\label{ampFlowUV}
\begin{aligned}
    \dot\psi + X^U\del_U(\psi) + X^V\del_V(\psi) &+ {1\over 2} \left(\partial_U( X^U) +\partial_V( X^V)\right) \\
     &+ \frac{r(U,V)-1}{2 r(U,V)^2}e^{-r(U,V)}\left(VX^U+UX^V\right)\psi = 0. 
\end{aligned}
\end{equation}

 As before, we will first solve for the velocity flow $X(s)$ using \eqref{eq:geodFlowXu} coupled with \eqref{unitspeed} and some initial $X(0)$. Once a satisfactory $X(s)$ has been obtained (which may need smoothing to reduce numerical noise), we will then resample it at a low resolution in order to have a faster computation when evolving initial data $\psi(0)$ or $\rho(0)$  using \eqref{ampFlowUV} or \eqref{densFlowUV} respectively. We used Mathematica NDSolve, which requires at least one spatial boundary condition during the flow evolution, which we took to be Dirichlet holding $X(s)=X(0)$ at the $U=U_{min}$ boundary. For $\psi(s)$ or $\rho(s)$, we use by default vanishing Dirichlet b.c.s at all four boundaries, $U=U_{min},U_{max}$ and $V=V_{min},V_{max}$.  

\subsection{Flow through the horizon and towards the singularity}

\begin{figure}
   \centering
       \includegraphics[width=\linewidth]{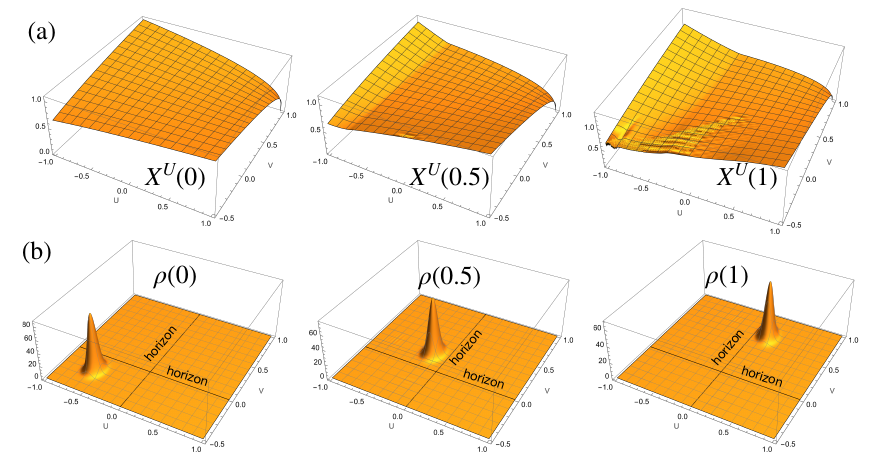}
\caption{Horizon-crossing plots for (a) evolved $X^U(s)$, (b) evolution of $\rho(s)$ starting at $s=0$ in region IV, moving by $s=0.5$ to region I and arriving by $s=1$ in region II where $U=0$ and $V=0$ horizons are marked in black.}
    \label{fig:horCross}
\end{figure}

A key benefit of working with Kruskal-Szekeres coordinates is that they allow for smooth crossing across the horizon, as already seen with the bunch of geodesics in Figure~\ref{fig:geodBunch}. We demonstrate this now for the flow of a bump across the horizons in our formalism. Here, the domains under consideration are $U\in[-1,1-\delta]$ and $V\in[-0.5,1-\delta]$ with $\delta=0.0000001$ that allows for all four regions with the $r=0$ singularity approached at the $(U_{max},V_{max})$ corner. For the initial vector field we take $X(0)=X_{sym}$ of (\ref{Xsym}) which does not obey $\div(X)=0$ but has the benefit of being widely defined. We then evolve this with the $U=-1$ boundary fixed, with results shown in Figure~\ref{fig:horCross}(a). This was computed at res=0.0001 and we see the emergence of a small amount of numerical noise at the $V=-0.5$ boundary. 
Note that it is perfectly possible to follow the strategy used before by solving \eqref{divfree} e.g. with boundary value  $1.1$ at the $U=-1$ boundary as discussed before but the resulting $X(0)$ has $X^U$ sharply peaked near the singularity (1,1) which then flips to a small value after a short time and indeed looks qualitatively the same as using $X_{sym}$.

For speed reasons, we next resample $X(s)$ with spacing $0.01$ in all three variables, which also reduces the visible noise at later times. Against this, we evolve a local density $\rho(0)$ Gaussian bump located in region IV with centre at $(-0.6,-0.2)$ and $\sigma=0.05$ to find that the bump crosses horizons smoothly during the evolution, moving from region IV to region I and then to region II as shown in Figure~\ref{fig:horCross}(b) (we show snapshots at $s=0,0.5,1$). We used res=0.0005 and as an integrity check on the numerical methods, the total integral of $\rho$ against the $\sqrt{-g}$  measure remains constant up to an error of $0.007\%$ over the range plotted. One can also choose an appropriate initial location of $\rho(0)$ to demonstrate the motion of a bump from region IV to region III and then to region I.

\begin{figure}
   \centering
       \includegraphics[width=\linewidth]{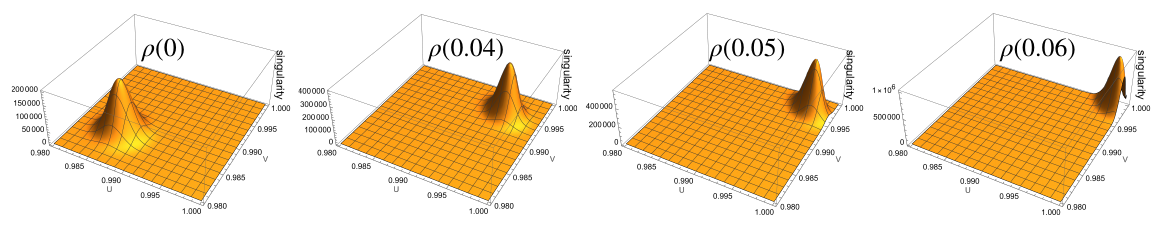}
\caption{Zoomed in approach to the singularity which meets our domain at (1,1). There is probability mass loss starting around $s=0.05$ as the bump spreads off stage on both sides of (1,1).}
    \label{fig:sing}
\end{figure}

It is possible by a fresh evolution from $\rho(1)$ to continue the flow inside region II until it exits the $V_{max}$ boundary of the domain, but nothing special happens (it would eventually hit the singularity off stage). Instead we show in Figure~\ref{fig:sing} a close-up approach to the singularity starting with a Gaussian bump centred at (0.986,0.986) with $\sigma=0.0015$. The domain is now $U,V\in [0.98,1-\delta]$ where we use the above $X(s)$ obtained by evolving $X(0)=X_{sym}$ now restricted to this smaller domain and resample this $X^U(s)$ with spacing $0.0005$ for $s$ and $0.0002$ for $U,V$ in order to speed up the $\rho(s)$ PDE without any visible change. For the $\rho(s)$ evolution we also set res=$\delta$ and drop the Dirichlet zero boundary condition at $U=U_{max}$ or $V=V_{max}$. The singularity is a hyperbola but meets our domain at the $(1,1)$ corner. One can see that the bump starts to flatten and some of its probability mass goes off stage on either side. It presumably follows the hyperbola but this would require a non-square domain to see clearly (instead we can see the cross-sections).

\subsection{Wave packet amplitude flow} 

As we explained in the Introduction, we can as well work with a complex wave function $\psi$ instead of density $\rho$. We demonstrate this using the same $X(s)$ that we used for a single bump in Section~\ref{sec:1bump}. As we saw in the Introduction, if $\psi$ has an initial polar decomposition, $\psi=\sqrt{\rho}e^{\imath\theta}$ then the phase $\theta$ simply evolves according to (\ref{angflow}) as convectively constant. For example, if we let $\psi(0)$ be a single Gaussian bump as before but with a phase factor then $|\psi|^2=\rho$ evolves exactly as before while $\theta$ imposes a wave-like structure on ${\rm Re}(\psi)$ and ${\rm Im}(\psi)$ which distorts but generally preserves its shape as $s$ increases. 

\begin{figure}
\centering
     \includegraphics[width=\linewidth]{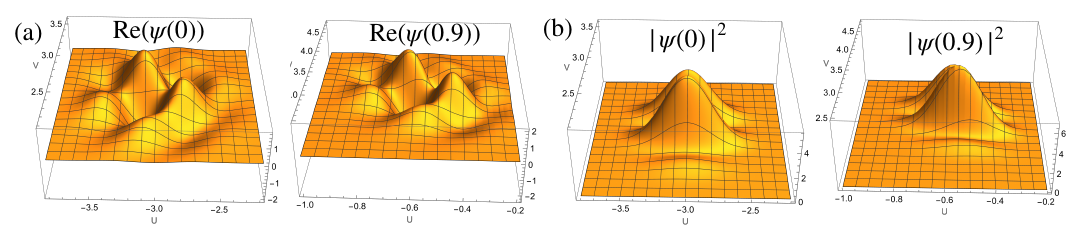}
\caption{Evolution of ${\mathrm Re}(\psi(s))$ and $\rho=|\psi(s)|^2$ in close-up for a dipole wave packet consisting a Gaussian bump and another one displaced by one standard deviation, with different phases. The base location evolves as in Section~\ref{sec:1bump} and the packet shrinks in width but maintains its  structure.}
    \label{fig:psidip}
\end{figure}

This is also the case for more complex wavefunctions, and here we demonstrate it for a dipole wave packet consisting of two Gaussian bumps displaced by one standard deviation and with different phases, namely
\[ \psi(0)(U,V)={1\over \sqrt{\sqrt{-g}Z}}\Big(e^{8\imath (U+V)} e^{-{(U-U_0)^2+ (V-V_0)^2\over 4\sigma^2}}+ e^{8\imath (U-V)}e^{-{(U-U_0)^2+ (V-(V_0-\sigma))^2\over 4\sigma^2}}\Big).\]
We have the same base location which evolves as for one bump in Section~\ref{sec:1bump}, while Figure~\ref{fig:psidip} shows the detail of ${\rm Re}(\psi)$ evolving, and there is a similar plot for ${\rm Im}(\psi)$. The difference now is that being the sum of different frequency modes, the density $\rho=|\psi|^2$ has interference patterns as for any wave packet, instead of being simply a Gaussian, as also shown. 
Note that in practice, we omitted the constant $Z$ normalisation factor, but a useful integrity test of the numerics is that the norm stays constant when computed correctly with the $\sqrt{-g}$ measure, up to numerical error of $0.7\%$ over the evolution computed at res=0.0005. By $s=0.9$ one can see small numerical noise in the form of ripples or striations of a type which will be studied in more detail in the next section for the case of colliding bumps. They have a wavelength is about 40 res and are fatter at res=0.001 but of similar amplitude.

\subsection{Collision of amplitude bumps vs density bumps} 

We now look at issues for constructing two $\psi$ wave packets vs two density  $\rho$ bumps that collide. Clearly, single geodesics can collide and each might be part of its own geodesic bunch which we can model as a density flow $\rho(s)$ and geodesic velocity flow $X(s)$ as we saw in Section~\ref{sec:1bump}. We can also extend this to wave packet flows $\psi(s)$ as we have just seen. But for two such flows to collide, we have a fundamental issue: what do we take for the velocity field, the one for one `particle' or the one for the other `particle'? In principle we could actually construct such statistical bunches and see what the composite $X_{stat}(s)$ looks like. But this would require a much more refined model than the crude estimate that we used and would only postpone the issue that, even for such modelling, when two particles on different trajectories cross, the physically relevant $X$ is multivalued. Interpolating to a smooth vector field without addressing this would depend critically on the interpolation method and not reflect the physics. One solution could be to let $X$ itself be quantum in some fashion so as to be able to take a superposition, or more precisely one might consider a tensor product system of some kind that includes both the amplitudes and the velocity fields of each particle. This may be looked at elsewhere.

Meanwhile, our approach within the present formalism (where $X(s)$ is a classical real vector field) is to interpolate the initial $X(0)$ for the two flows and then let this single vector field evolve with $s$ as a geodesic velocity field. The physical interpretation of this is unclear for the reasons above (basically two particles or two particle bunches is not how we think of a geodesic flow), but we can take any initial $X(0)$, so we are free to do this. The down-side, however, is that as the system evolves, if at some point two wave functions or densities approximately coincide, then they must both continue to evolve approximately the same, as there is only one (shared)  velocity vector field $X$. What this means is that we can, and shall, model two particle `bumps' meeting but they will tend to move together thereafter rather than to actually cross each other. It is not impossible that they cross, e.g., if one has a very different shape from the other, but exhibiting this would appear to take a much higher resolution than available.

\subsubsection{Evolution of the interpolated velocity field and smoothing}\label{sec2bumpsX}

\begin{figure}
   \centering
      \includegraphics[width=\linewidth]{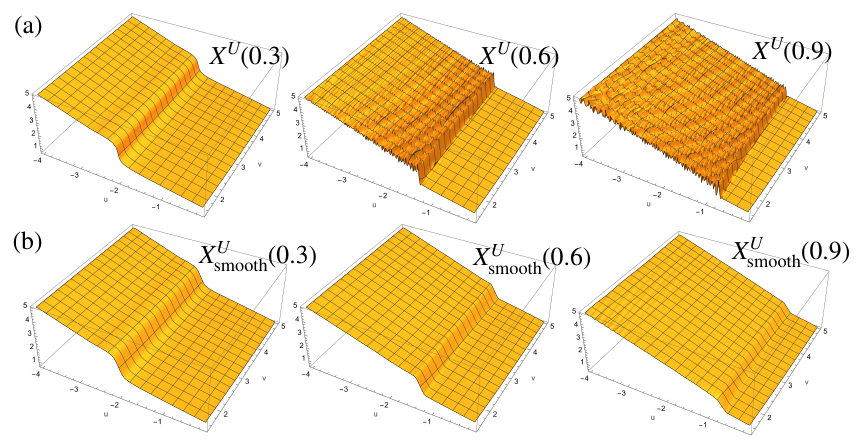}
\caption{Plots for (a)  $X(s)$ needed for the 2-bump collision computed at res=0.003 and (b) the same after smoothing.}  
\label{2bumpsX}
\end{figure}

To be concrete, we let `particle 1' be a positive $\psi$-bump centred at $(-3.5,2.5)$ with $X_1(0)$ the same type as used in Figure~\ref{fig:Xstat}. Let `particle 2' be a negative $\psi$-bump centred at $(-2.6, 2.2)$ and $X_2(0)$ given by constant value $3$ on $U=U_{min}$, i.e. with very different velocities. We then tanth-interpolate these with
\[ X^U(0)=\frac{1}{2} (1-\tanh (3 (U+3))) X^U_1(U,V) +\frac{1}{2} (1+\tanh (3 (U+3))) X^U_2(U,V)\]
as our initial velocity field. We set $X^V$ from $X^U$ so as to be unit speed. We then let  $X(0)$ evolve using res=0.003 with results as shown in Figure~\ref{2bumpsX}. We see that the transition zone between the two regions moves towards the $U=0$ horizon. Here $V$ ranges from $V_{min}=1.5$ to $V_{max}=5$. However, we also see that as $s$ increases, the solution rapidly develops numerical noise/ripples. The ripples also grow in amplitude as we approach the $V=0$ horizon and we have set $V_{min}=1.5$ to reduce their impact. They are not, however, caused by the edge of the domain (where we do not impose any condition) since we can set $V_{min}=0.5$ and still have substantially the same cross-section at a given $V$ in the interior. Our approach to these ripples is that they are numerical artefacts and should be smoothed out as shown in the second row. The smoothing is done for each $s$ by evaluating $X^U$ on a uniform $(U,V)$-grid, to each 2D dataset we apply a Gaussian convolution of a fixed small radius and then re-interpolate this filtered data. The result is shown as $X^U_{smooth}(s)$.

\begin{figure}
\centering
\includegraphics[width=\textwidth]{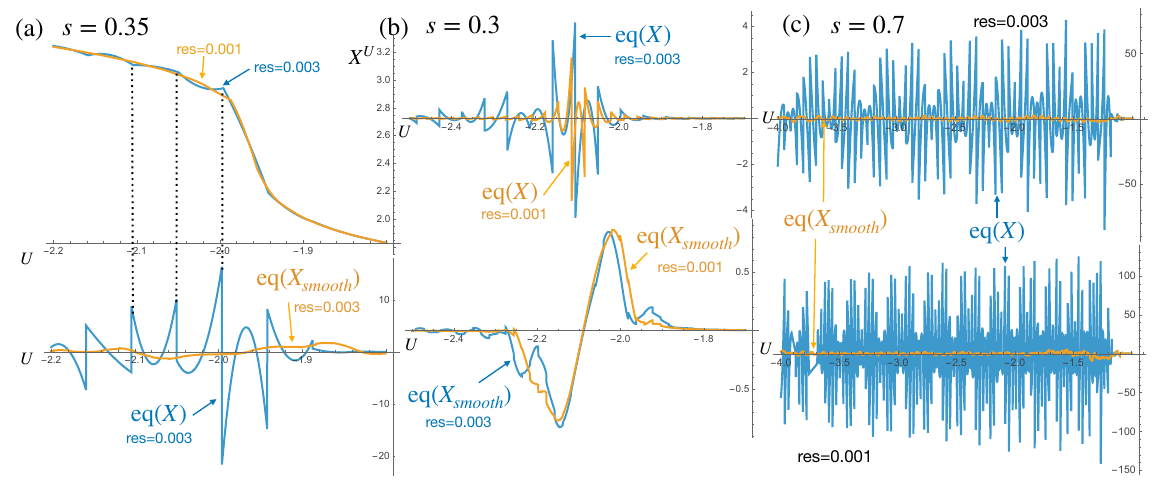}
\caption{Cross sections at  $V=3.25$ showing improved accuracy changing resolution from res=0.003 to res=0.001 in (a) at $s=0.35$. The residual spikes at the segment boundaries in the piecewise solution but for smaller $s$, shown in (b) for $s=3$, reduces at finer resolution. We also see the effect of smoothing on the slope at the transition region. But at late times, shown in (c) for $s=0.7$, the residual spikes more for finer resolution.}  \label{Xripples}
\end{figure}

Figure~\ref{Xripples} clarifies the numerical nature of these ripples. From part (a) we can see  at $s=0.35$ that $X^U$ at res=0.003 has cusps where the derivatives do not match, due to the solution being constructed piecewise in segments of around 15 res, while at finer res=0.001, these ripples are barely visible. Below the solution we plot the residual ${\rm eq}(X)$ as the left hand side of (\ref{velocityflow}), which should be zero for a solution. This has spikes at the segment boundaries due to failure of the derivatives and hence of ${\rm eq}(X)$ to be properly defined there. Mathematica does not complain, but evidently the solution aims to be continuous but not differentiable. On the other hand, ${\rm eq}(X_{smooth})$ is more reasonable and shows that we have a fair solution after smoothing. Note that we do not study the resampled $X_{smooth}$ which has further (but comparable) numerical artefacts from the resampling. In part (b) we return  to $s=0.3$ and check that the finer resolution res=0.001 indeed has a smaller residual. However,  ${\rm eq}(X_{smooth})$ does not decrease but tends to a specific sideways S shape. This is because the smoothing also smooths the sharp drop in $X^U$ so that the smoothed $X_{smooth}$ does not quite obey the velocity equation in the transition region, but this is a known and fixable error (we could for example apply the smoothing only away from the transition). Aside from this complication, we see that we solve the velocity flow equation with greater accuracy for finer resolution.

This applies for smaller $s$, but by the time $s=0.7$ in part (c) of the figure, we are in a different regime where the ripples in the solution are similar in amplitude (but smaller in wavelength) for finer res=0.001. In fact the spikes in ${\rm eq}(X)$ actually increase for finer resolution, while  ${\rm eq}(X_{smooth})$ does not change much with resolution. It is much smaller than before smoothing but still not small compared to the size of $X^U$, so we only have a moderately good solution even after smoothing and cannot improve it with finer resolution. One could smooth further (with a larger smoothing radius) but this would cause more error at the transition region. An approach for better solutions would be only running the evolution in small enough intervals of $s$ so that the chaotic behaviour at later time does not set in, smoothing, and then using this as the initial state for another such interval, etc.  Hence, a smooth solution more strictly obeying (\ref{velocityflow}) for all times could be constructed in principle using less smoothing over several intervals, but from a distance it is not expected to look very different from the second row of Figure~\ref{2bumpsX}, just a slightly steeper transition as this was blunted by the smoothing.

\subsubsection{Evolution of amplitude $\psi$ bumps to a dipole vs  evolution of density $\rho$ bumps}

In this section, we solve the $\psi$ evolution (\ref{ampflow}) with respect to the smoothed $X(s)$ obtained in the preceding section, except that we resample it with $0.02$ intervals in $s$ and $0.1$ intervals in $U,V$. This is purely to speed up the numerical solutions in this section, without changing $X^U$ too much. It consists of a step in $X^U$ moving  towards the $U=0$ horizon and looks identical to Figure~\ref{2bumpsX}(b) but without visible ripples by the time we have done the resampling. 

\begin{figure}
 \[ \includegraphics[width=\linewidth]{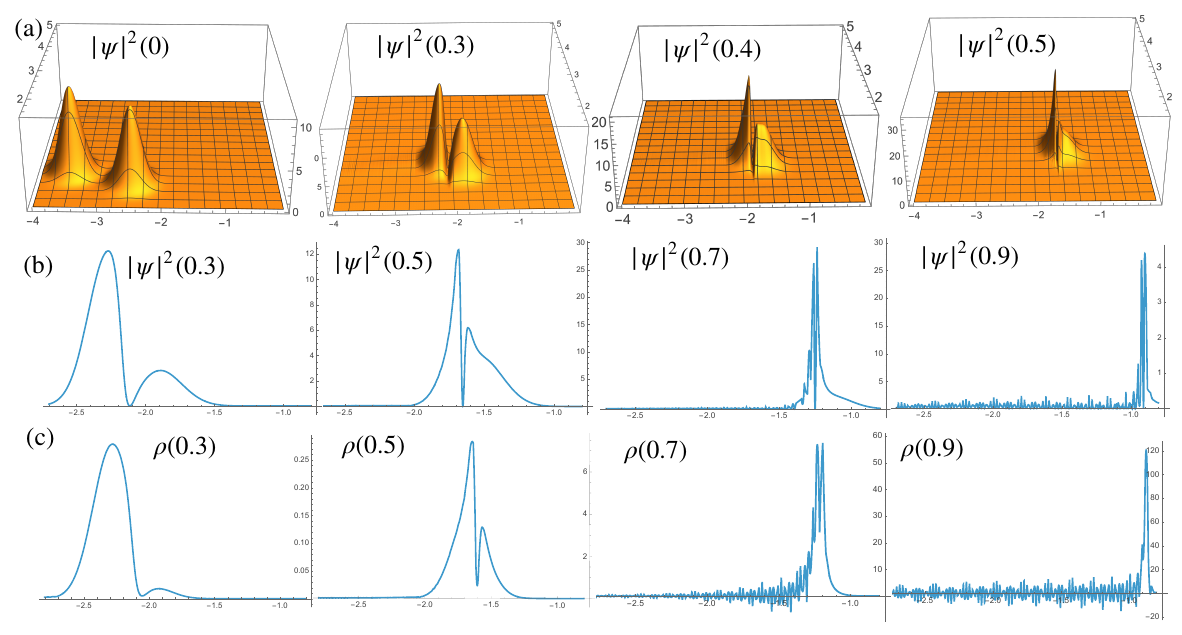}\]
     \caption{(a) Two `particle' bumps evolving with respect to $X(s)$ from Section~\ref{sec2bumpsX}. Up to $s=0.5$, the displayed case of evolving $|\psi|^2$ where one of the $\psi$-bumps is negative is qualitatively similar to  case of two Gaussian bumps and evolving $\rho(s)$; (b) Cross-section of $|\psi|^2$ at $V=2.9$ for opposite sign initial $\psi$-bumps colliding to  a dipole; (c) Cross-section at $V=3.3$ of initial $\rho$-bumps colliding to another bump. The trailing peaks or ripples in the two cases are numerical noise.}    \label{2bumps_cross}
\end{figure}

Relative to this velocity field, we take the initial $\psi(0)$ to be a superposition of two bumps of different phases (we chose to keep $\psi$ real for simplicity, so one bump is positive and the other is negative). The bump centres are located such that classical geodesics with velocity starting with the value of $X(0)$ at the two locations would meet and cross-over near $s=0.65$. Interestingly, this does not happen --  the two `particles' indeed move towards each other but when they collide they turn into and continue to evolve as a real dipole. This is shown in Figure~\ref{2bumps_cross}(a) for the run-up and in cross-section at $V=2.9$ in Figure~\ref{2bumps_cross}(b) from $s=0.3$ onwards. As an integrity check \eqref{integrity}, with $\rho=|\psi|^2$ here, we find during the evolution that it remains constant up to  $2.8\%$ over the full range of $s$ discussed. Note that the positive and negative $\psi$-bumps can never `catch up' and still maintain their shape since to do so would drastically reduce the norm due to overlap with opposite sign. The cross-section images show some ambient trailing numerical noise at the used res=0.0005, which should be ignored. 

By contrast, we show the $\rho$-evolution of two density bumps where each bump is the $|\psi_i|^2(0)$ of the previous wave-function bumps. (This is equivalent to two wave function bumps that are of the same sign or phase instead of opposite.) Note that this initial $\rho(0)=|\psi_1|^2(0)+ |\psi_2|^2(0)$ is not exactly the same as $|\psi_1+\psi_2|^2(0)$ because the initial bumps overlap slightly, which is just as well since  $\rho$ and $|\psi|^2$ obey the same differential equation so they could not evolve differently if they had exactly the same initial value. Since the densities also evolve additively, the result looks the same as one density bump catching up with the other bump, approximately the same as in Figure~\ref{2bumps_cross}(a), but this time we see in the cross-sections in Figure~\ref{2bumps_cross}(c) that they merge into a single bump. Note that the bumps at $s=0.3$ are of similar size but do not look like that because the second one is centred even further below the $V=3.3$ cross section. We used the same resolution res=0.0005 but the evolution is more noisy. Note that the trailing ripples can only be numerical artefacts (and are indeed bigger for coarser resolution) because $\rho$ cannot become negative in the evolution. The total mass with respect to the measure again remains constant during the evolution to $4.4\%$, as an integrity check \eqref{integrity}  given the amount of noise.

Note that two density bumps that do not quite collide might still appear as like a dipole with respect to the density profile, to some extent. By contrast, the wave function case cannot end up with the positive and negative bumps exactly on top as explained, resolved in our case by the dipole outcome. We do not know fully how the difference between these two cases  translates to a new physical prediction because wave functions are not normally taken on all of spacetime and quantum mechanics is not normally with respect to some kind of collective proper time $s$ as it is here. However, these issues are matters of interpretation and if they can be overcome then we see a route to testing our hypothesis of an underlying wave function amplitude as opposed to the evolution being that of densities.

\section{Geodesics on quantum phase space}\label{sec:kg}

So far, we have been doing only classical geometry around a black-hole, but using the language of quantum geodesic flows, except that we also considered the possibility that $\rho=|\psi|^2$ where $\psi$ obeys (\ref{ampflow}). This equation, while first order, is not, however, Schr\"odinger's equation. For something like the latter (but for evolution with respect to $s$), one should look instead at quantum geodesic flows on the Heisenberg algebra\cite{BegMa:geo} or in general on the algebra $\CD(M)$ of differential operators\cite{BegMa:qm}. In this section, we revisit the black-hole case in more detail than in \cite{BegMa:qm} by using Kruskal-Szekeres coordinates \(U,V,\theta,\phi\) as recalled in Section~\ref{sec:pre}.

\subsection{Heisenberg picture evolution in Kruskal-Szekeres coordinates}
We first compute the contracted Christoffel symbols $\Gamma^\mu$, in units of $r_s=1$, using \eqref{ChristoffelUV} to find,
\begin{equation}
\Gamma^U = -\frac{U}{r(U, V)},\quad
\Gamma^V = -\frac{V}{r(U, V)},\quad
\Gamma^\theta = -\frac{\cot\theta}{r(U, V)^2},\quad
\Gamma^\phi = 0.
\end{equation}
Now, using the $\Gamma^\mu$ expressions above, we can easily find the $4$-momenta from \eqref{momenta},
\begin{equation}
\begin{aligned}
    p_U &= \frac{1}{r(U, V)^2}e^{-r(U,V)}(\lambda V - 2m \frac{\extd V}{\extd s}),\quad p_V = \frac{1}{r(U, V)^2}e^{-r(U,V)}(\lambda U - 2m \frac{\extd V}{\extd s}),\\
    p_\theta &= -\frac{\lambda}{2} \cot\theta + m\,r(U, V)^2\frac{\extd \theta}{\extd s},\quad p_\phi = m\, r(U, V)^2 \sin^2\theta\,\frac{\extd \phi}{\extd s}.
\end{aligned}
\end{equation}
Similarly, the flow equation for these $4$-momenta in \eqref{momentaFlow} comes out to be,
\begin{equation}
\begin{aligned}
    -m \frac{dp_U}{ds} &= \frac{\lambda}{2r(U,V)}p_U + \frac{V}{2r(U,V)}(1+r(U,V))p_Vp_U \\
    &\qquad\qquad\qquad+ \frac{\lambda V}{2r(U,V)^3}e^{-r(U,V)}(U p_U + V p_V) + \frac{V}{r(U,V)^4}e^{-r(U,V)}p_{sph}^2,\\
    -m \frac{dp_V}{ds} &= \frac{\lambda}{2r(U,V)}p_V + \frac{U}{2r(U,V)}(1+r(U,V))p_Vp_U \\
    &\qquad\qquad\qquad+ \frac{\lambda U}{2r(U,V)^3}e^{-r(U,V)}(U p_U + V p_V) + \frac{U}{r(U,V)^4}e^{-r(U,V)}p_{sph}^2,\\
    m \frac{dp_\theta}{ds} &= \frac{\lambda}{2r(U,V)^2\sin^2\theta}p_\theta + \frac{\cot\theta}{r(U,V)^2\sin^2\theta}p_\phi^2,\\
    m \frac{dp_\phi}{ds} &= 0,
\end{aligned}
\end{equation}
where the spherical momentum $p_{sph}^2$ is defined as
\begin{align}
    p_{sph}^2 = p_\theta^2 + \lambda \cot\theta\,p_\theta + \frac{p_\phi^2}{\sin^2\theta}. 
\end{align}
There is also the total momentum given by,
\begin{equation}\label{laplacianUV}
    -p_{tot}^2 = \frac{\lambda}{r(U,V)}(U p_U + V p_V) - r(U,V)e^{r(U,V)}p_Up_V + \frac{1}{r(U,V)^2}p_{sph}^2.
\end{equation}
The $p_\mu$ will be represented as evolving operators, but if we take them as classical real numbers and set $\lambda=0$ then setting $p_{tot}^2$ to a constant gives the equations of a single geodesic in first order form while $p_{sph}^2$ is the conserved angular momentum.  The operator formulae can be used for Ehrenfest theorem applications, see \cite{BegMa:qm}, but we have mainly recalled them as they provide the interpretation of $\extd s$ as proper time. Instead, we now focus on the corresponding Schr\"odinger picture. 

\subsection{Klein-Gordon flow as pseudo-quantum mechanics}
We are interested in the evolution of radial wave functions $\phi=\phi(U,V)$ under the Klein-Gordon flow. The Klein-Gordon operator here is given by the Schr\"odinger representation where $p_\mu$ act as $\lambda\frac{\del}{\del x^\mu}$ for $x^\mu \in \{U,V\}$ in \eqref{laplacianUV} and $\lambda=-\mathrm{i}\hbar$ and 
\begin{equation}
    -\lambda \frac{\extd \phi}{\extd s} = \frac{-p_{tot}^2}{2m}\phi
\end{equation}
recovers (\ref{KGflow}) explicitly as
\begin{equation} \label{KG1}
    -\mathrm{i}\frac{\del \phi}{\del s} = \frac{\hbar}{2m}\left(\frac{1}{r(U,V)}(U\frac{\del\phi}{\del U}+V\frac{\del \phi}{\del V}) - r(U,V)e^{r{(U,V)}}\frac{\del^2\phi}{\del U\del V}\right).
\end{equation}
Unlike in Schwarzschild coordinates $(t,r)$, there is no explicit time direction here, but the energy conservation still holds. We make use of this by demanding our wave function $\phi$ is an eigenfunction of the Killing vector field $\del_t$ \eqref{deltKilling} of the form $e^{\mathrm{i}pt}$, which using \eqref{UVrelations} can be written as, 
\begin{equation}\label{n-psi}
    (-{U\over V})^{-{\rm i} p} \psi(s,z)
\end{equation}
in $U,V$ coordinates with $z:=UV$ (we then take this form in other regions). Here, $p$ determines the conserved energy, while $\psi$ depends only on $z$ with $z<0$ in regions I, III and $z>0$ in regions II, IV. The $U=0$ horizon is at $z=0$. With this form of fields, the Klein-Gordan flow \eqref{KG1} becomes,
\begin{equation}
    -\mathrm{i}\frac{\del \psi}{\del s} = \frac{\hbar}{2m}\left(\frac{2z}{r(z)}\psi^\prime - r(z)e^{r{(z)}}\big(\frac{p^2}{z}\psi+\psi^\prime+z\psi^{\prime\prime}\big)\right),\label{kgqm}
\end{equation}
where prime means derivative with respect to $z$ and $r(z)=1+W(-z/e)$. We also use the Jacobean for the change of coordinates from $(U,V)$ to $(z,w)$ where $w:=U/V$, and get the effective measure from \eqref{measureUV} as
\begin{align}
{ r(z) e^{-r(z)} \over |w|}\extd z\extd w
\end{align}
for the $L^2(M)$-norm on the original fields \eqref{n-psi}.
Following the strategy of \cite{BegMa:qm} (which is also broadly the strategy in the derivation of quantum mechanics) we then have a kind of `pseudo-quantum mechanics' where we can complete with respect to the $L^2$-inner product on $\psi(z)$ now with measure $r(z) e^{-r(z)}\extd z$  and use the methods of conventional quantum mechanics in one variable $z$ and `time' $s$. One can do the same when there are angular variables, but we focus on the radial theory. Note that in spite of structural parallels, we are not doing actual quantum mechanics in any usual sense due to evolution being with respect to the collective proper time $s$ rather than a given laboratory time.

\begin{figure}
         \includegraphics[width=\linewidth]{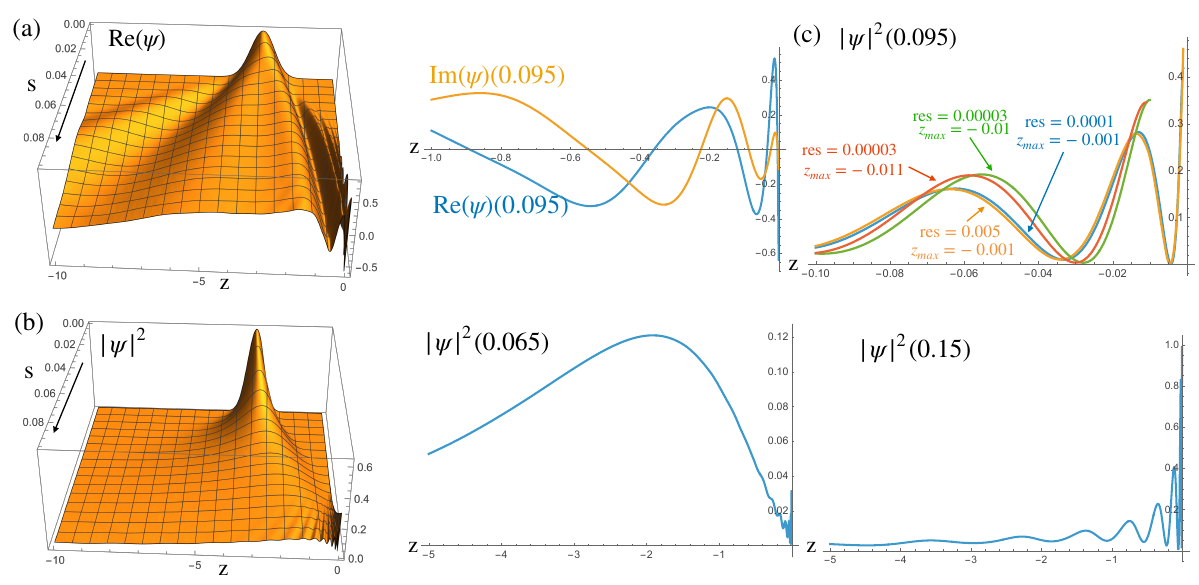}
       \caption{Horizon modes when the area of disturbance from a dissipating Gaussian approaches the $U=0$ horizon. (a) shows the evolution of ${\rm Re}(\psi(s,z))$ while (b) that of $|\psi(s,z)|^2$. We also show crosss-sections at when the modes start to appear at around $s=0.06$ and later at $s=0.15$, by which time the horizon modes have replaced the Gaussian bump. (c) Checks robustness under change of resolution and change of the right boundary $z_{max}$.}  \label{horiz}
\end{figure}

As a concrete example, we set $p=-1$ with boundary conditions $\psi(s,z_{min})=0$ at large negative $z_{min}$ (i.e, far from the $U=0$ event horizon). For convenience we show results for $\hbar=m=1$ and we start off with an initial Gaussian wave-function centered around $z=-3$ with $\sigma=0.5$. We set $z_{min}=-30$ to be far from any area of disturbance for the range of $s$ considered and we set $z_{max}=-0.001$ so as to work just above the horizon in region I. Results using Mathematica with res=0.0001 are shown in Figure~\ref{horiz} with plots of (a) the real and (in cross-section) imaginary parts of $\psi$ and (b) the density $|\psi|^2$.  We see that  as $s$ increases, the Gaussian peak moves towards the horizon and dissipates in a wave-like manner as expected but when the area of disturbance approaches the $U=0$ horizon at around $s=0.065$,  it creates `horizon modes' that increase with $s$ until by $s=0.15$ the original Gaussian has been `eaten' by the black-hole and turned wholly into horizon modes. This reproduces and confirms this novel phenomenon found in \cite{BegMa:qm} in Schwarzschild $(t,r)$ coordinates. We have also confirmed as in \cite{BegMa:qm} that the wave function entropy (i.e. that of $|\psi|^2$ as a probability density) increases throughout this process while the expected value $\langle z\rangle$ actually becomes more negative, i.e. heads away form the horizon due to the horizon modes. Increasing the energy of the system, i.e. increasing $p$, makes the horizon modes  appear at smaller $s$. As an integrity check we also verified that $|\psi|^2$ with the above measure is constant during the evolution up to around $0.06\%$.

Using our Kruskal-Szekeres coordinates, we can also look more carefully at these horizon modes as we approach the horizon. Here,  part (c) of Figure~\ref{horiz}  checks the numerical robustness under changes of resolution. In fact there is no visible change as we increase res (corresponding to a coarser resolution) until ${\rm res}>|z_{max}|$, i.e. if we do not try to get  too close to the horizon. This is shown for $z_{max}=-0.001$ and res=0.005 vs res=0.0001. Changing $z_{max}$, while it  does not affect the horizon modes far from the horizon (the curves tend to each other at large negative $z$), does introduce significant differences closer to the horizon in both phase and wavelength. We found that this dependence is muted if $|z_{max}|>1000\ {\rm res}$ or so. We show results for ${\rm res}=0.00003$ and $z_{max}=-0.01$ vs $z_{max}=-0.011$. We also looked at the residual (the apparent failure of the equation to be solved) but here, unlike Section~\ref{sec2bumpsX},  the residual at a fixed $z$ does appear to tend to zero for small res (up to other forms of numerical noise such as from machine precision). It does not present as small for the resolutions plotted but as previously explained, this has to with solutions   being  constructed in piecewise segments with non-matching derivatives at the segment boundaries. On the other hand, for a fixed resolution, the residual increases as we approach $z=0$, confirming that results are unreliable if we try to get too close to the horizon. This is attributable to the $1/z$ term in (\ref{kgqm}) which causes increasingly fast (but apparently bounded) oscillations as we approach the horizon. 

\begin{figure}
        \includegraphics[width=\linewidth]{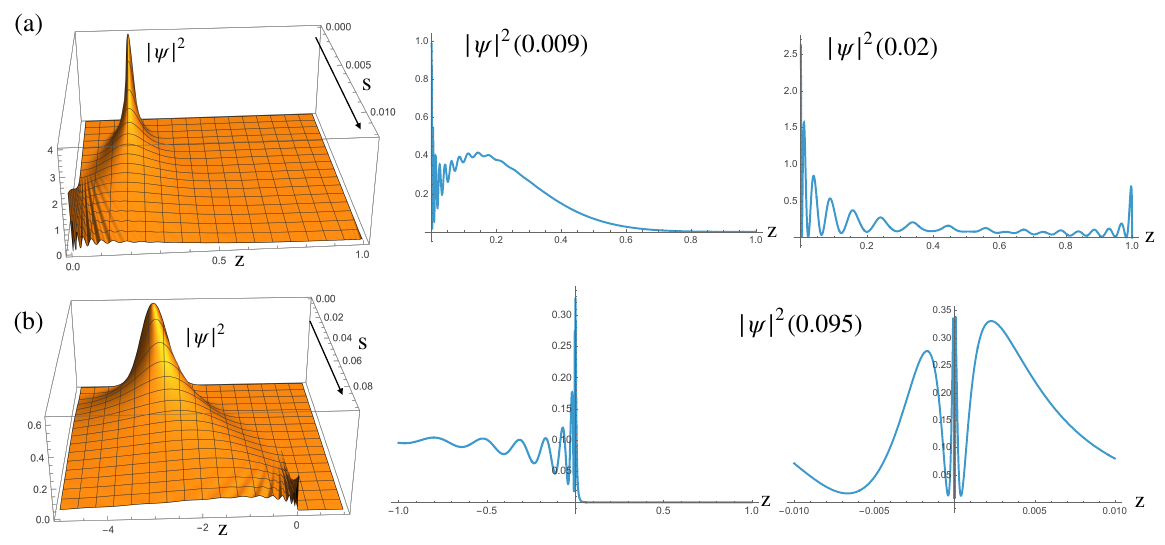}
       \caption{(a) Gaussian bump inside the event horizon creates horizon modes like a mirror of the one outside. (b) Purported  solution crossing the horizon for discussion purposes.} \label{horiz_inner}
\end{figure}

Finally, given that the Kruskal-Szekeres coorindates are also fine at $z\ge 0$, we could try to solve the Schroedinger-like equation inside the black-hole.  Figure~\ref{horiz_inner}(a) shows similar behaviour for a Gaussian bump inside but with horizon modes generated on the inside of the horizon. We used the same base resolution res$=0.0001$ but now $z_{min}=0.001$ and $z_{maz}=1$. The Gaussian peak actually moves slightly towards the horizon as it gets absorbed there. One has similar issues in that now $z_{min}$ cannot be too small for reliable numerics due to the increasingly fast oscillations approaching the horizon from $z>0$, and some dependence on its value even at high resolution. We did not plot it but checked that $\langle z\rangle$ increases throughout, i.e. moves away from the horizon and towards the centre of the black-hole. Likewise, the state entropy increases till about $s=0.017$ and then starts to slightly decrease by $s=0.02$. We attribute the latter behaviour to unphysical modes at $z=1$ which start to be visible at $s=0.014$ when the rate of increase of entropy slows down. These modes, as in the last plot of Figure~\ref{horiz_inner}(a) at $s=0.02$, are not physical as they are artefacts generated by reflection on the $\psi(z_{max})=0$ boundary condition artificially imposed there. We have deliberately shown the plots to $s=0.02$ to exhibit these modes but the physical process should only be taken to before the area of disturbance visibly hits the $z=1$ boundary. 

Meanwhile, Figure~\ref{horiz_inner}(b) purports to follow our original Gaussian bump in the exterior but now also looking inside the black-hole by setting $z_{max}=1$. The second cross-section is a close-up near the horizon. One can blow up by a factor of 10 and it looks similar (but more mirror-symmetric). These plots should be viewed with caution because while Mathematica claims to solve the evolution without any problem, we  already know from our analysis for the left (and right) sides that the numerical evolution is not robust if we extend the region of integration from either side too close to the horizon. Hence, while the plots would seem to suggest that the horizon modes excite mirror image modes inside the horizon, this could be an effect of the discretization which cuts off the infinitely fast oscillations as we approach $z=0$ and effectively smooths out the plot at the resolution scale.  

What this numerical study implies for the continuum theory is that solutions to the Klein-Gordon flow in the fixed energy sector (what we called pseudo-quantum mechanics) are not differentiable at the horizon due to horizon modes having increasingly small wavelengths approaching it. These modes arise from a wave packet of (originally a Gaussian peak in our case) approaching the horizon, where in a finite time $s$ the original field is absorbed and replaced by horizon modes as in \cite{BegMa:qm}. These modes also appear to excite mirror image modes on the other side of the horizon but this could be a numerical artefact since we cannot numerically evolve across the horizon without blurring over the infinitely small wavelengths on either side. On the other hand, however, spacetime is likely {\em not} a continuum and there may be some kind of discreteness/noncommutativity at the Planck scale, due to quantum gravity effects\cite{Ma:pla, DFR, MaRue}. Judging from the calculations above, this could smooth out the infinite frequencies at $z=0$, basically cutting these off at the Planck scale with the information from the swallowed wave packet eventually residing in a kind of `skin' at the horizon. Such a `skin' would be in the same ball-park as in an older (wave-operator) approach to noncommutative black-holes in \cite{Ma:alm}. These remarks may also be relevant to `atomic' states in \cite{BegMa:qm} defined as bounded stationary modes (i.e. actual solutions of the Klein-Gordon equation) of the evolution (\ref{kgqm}), where it would be of interest to see if these can be followed through the horizon and also if there are atomic modes inside the black-hole. We turn to this next.

\subsection{Atomic modes for the Klein-Gordon wave operator}
Another approach to solving the geodesic flow in this phase-space system is to consider eigenmodes $E_{KG}$ of the Klein-Gordon operator \eqref{laplacianUV},
\begin{align}
	{-p_{tot}^2 \over 2m}\phi  = E_{KG} \phi,
\end{align}
which is akin to solving the corresponding `time-independent' Schr\"odinger equation. Since this system admits another Killing vector field $\del_\phi$ (for angular momentum conservation) apart from $\del_t$ as we saw previously, we make use of such eigenfunctions to write down $\phi(U,V,\theta,\phi)$ in a separable form as
\begin{equation}
	(-{U\over V})^{-{\rm i}p} \psi(s,z) e^{{\rm i}n\phi} F(\theta).
\end{equation}
This leads, following routine calculation involving separation of variables, to
\begin{equation} \label{radKGmode}
{2mE_{KG} \over \hbar^2}\psi + \left(\frac{2z}{r(z)}\psi^\prime - r(z)e^{r{(z)}}\big({p^2 \over z}\psi+\psi^\prime+z\psi^{\prime\prime}\big)\right) = {l(l+1) \over r(z)^2}\psi,
\end{equation}
for some constant parameter $l$ along with the Laplace equation on $\R^3$, i.e.
\begin{equation}\label{lapR3}
\left( l(l+1) - {n^2 \over \sin^2\theta} \right)F + \cot(\theta) F' + F'' = 0,
\end{equation}
where prime on $F$ refers to a derivative with respect to $\theta$. The latter is solved by Legendre polynomials $P^n_l(\theta)$ with $n=0,\ldots,l$ as part of the well-known spherical harmonics $Y_{l,n}(\theta,\psi):=P^n_l(\theta)e^{\mathrm{i}n\phi}$. It has been shown in \cite{BegMa:qm} that this system has similarities with quantum mechanical modes of a Hydrogen atom.  Here we work with parameters $\hbar=m=-p=1$ and solve \eqref{radKGmode} with different types of modes separated by the critical value $E_{KG}=-{p^2 \over 2m}=-{1\over 2}$ in our case. This is the same as in \cite{BegMa:qm} but it should be noted that the solutions on either side do not look too different in practice but belong to different families. Increasing $p$ makes them more and more oscillatory for large negative $z$, while all modes are fractally oscillatory at the horizon. We study solutions for the spherically symmetric case, i.e.  $l=0$, and comment on the effects of higher $l$ values for the two kinds of solutions below.

\begin{figure}
        \includegraphics[width=\linewidth]{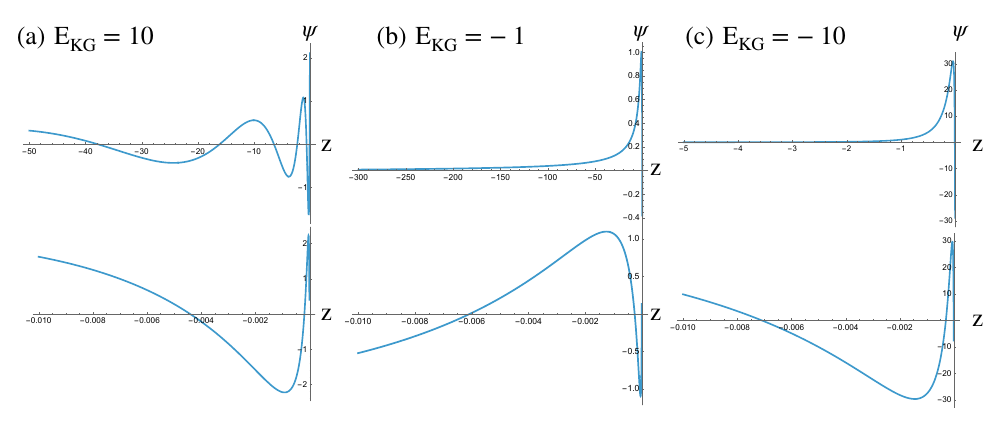}
       \caption{Klein-Gordon modes outside the horizon for three values of $E_{KG}$ on the top row and their close-ups near the horizon on the bottom row.} \label{outerKGmodes}
\end{figure}

Solutions outside the horizon are shown in Figure \ref{outerKGmodes} computed with res=0.00001 and $z_{min}$ set to be a large negative value (depending on the mode under consideration) to be far from the horizon, while $z_{max}=-{\rm res}$ is set right above the horizon. The plots do not look much different for larger $|z_{max}|$ other than missing detail near the horizon. We used a mixed boundary condition at $z=-1$, i.e. sufficiently far from the horizon, which as in  \cite{BegMa:qm} we fine-tune to get the solutions (b) and (c) that would otherwise have diverged. The solution (a) is an oscillatory mode typical of a large positive $E_{KG}=10$  with decreasing amplitude as we move away from horizon for increasingly negative $z$ but not decaying fast enough to be have finite norm. As we decrease the value of $E_{KG}$ towards the critical value, we see oscillations increasing in wavelength (and decreasing amplitude) far from the horizon. By contrast, (c) shows a solution typical for large negative  $E_{KG}=-10$, which rapidly approaches zero away from the horizon and has finite norm. Closer to the critical value,  $E_{KG}=-1$ in (b) requires a bigger right boundary at $z_{min}=-300$ but is otherwise similar. It comes down less sharply and as we approach $E_{KG}=-0.5$, one has to keep pushing the right boundary to a very large negative $z_{min}$ for a similar picture. For all three solutions, we see  fractal like behaviour as in \cite{BegMa:qm} near the horizon in the closeups directly below. Additionally, we find that increasing the value of the angular parameter $l$ results in similar modes except that they now start to diverge with increasingly higher values near the horizon. We also checked that their residual in the sense of the failure of (\ref{outerKGmodes}) is tolerable as long as we do not get too close to the horizon. These results are qualitatively the same and hence confirm those in Schwarzschild coordinates in \cite{BegMa:qm}.

\begin{figure}
        \includegraphics[width=\linewidth]{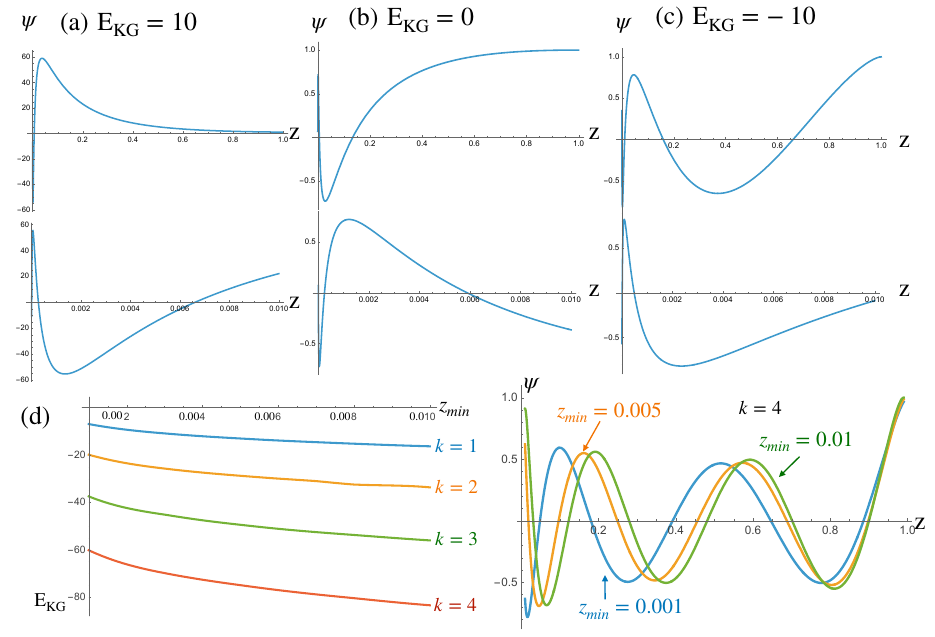}
       \caption{Klein-Gordon modes inside the horizon (a)-(c) for three different values of $E_{KG}$ in the first row, along-with their close-ups near horizon shown directly below in the second row. (d) $z_{min}$-dependent discretisation of the eigenvalues $E_{KG}$ if we impose $\psi'(z_{min})=0$ as a model of quantum gravity corrections} \label{innerKGmodes}
\end{figure}

Next, since we used Kruskal-Szekeres coordinates, we can now equally well solve for similar modes inside the black-hole. Solutions are in  Figure~\ref{innerKGmodes}(a)-(c), again with res=0.00001 but now setting $z_{min}$=res and $z_{max}$=1. If we employ a mixed boundary condition  at say $z=0.5$ in line with the approach outside the black-hole, solutions typically diverge at $z=1$ but can be fine-tuned to any value to remove the divergence. Given this behaviour, we directly focussed on such solutions by setting $\psi(1)=1$ and $\psi'(1)=0$ (but one is free to choose other values at $z=1$). This approach also avoids artefacts appearing in the residuals in the middle of the region. The qualitative nature of the solutions found is a mirror of the modes outside the horizon, now with large positive  $E_{KG}$ as in (a) yielding bounded modes that (after the fast oscillations near the horizon) decay as we approach $z=1$. Large negative values as in (c) produces oscillatory modes right up to the  $z=1$ singularity at the centre of the black-hole. There is again a soft $E_{KG}=-0.5$ with (b) near to this value. In all three cases, we again have fractal-like oscillations approaching the horizon as shown in the close-ups directly below. Furthermore, we find that increasing the parameter $l$ results in appearance of more oscillation cycles before reaching the singularity, irrespective of the value of $E_{KG}$. 

Finally, we note that outside the horizon as in \cite{BegMa:qm} and inside it, there is no quantisation of the values of $E_{KG}$, which can be attributed to the lack of any boundary condition at the horizon itself. Mathematica was not able to solve for modes across the horizon but supposing we could do it, we might aim for $\psi'(0)=0$ due to the symmetric nature of solutions expected there (cf the last plot of Figure~\ref{horiz_inner}(b)). Hence a regularisation scheme to eliminate the highest frequency oscillations too close to the horizon would be to solve separately on either side of the horizon with  $\psi'(z_{max})=0$ for the patch outside and $\psi'(z_{min})=0$ for the patch inside, where $|z_{max}|=z_{min}$ so that we are keeping symmetry on either side of the horizon. All of the detail closer to the horizon is then missing, with the smallest value $|z_{max}|=z_{min}={\rm res}$ losing the least (but also the most unreliable) information. Focussing on the black-hole interior, we used the  $\psi'(z_{min})=0$ boundary condition to replace the previous $\psi'(z_{max})=0$ boundary condition but then we searched (by computer) for $E_{KG}\le -0.5$ such that we still land on $\psi'(z_{max})=0$ anyway. This indeed results in a discrete spectrum with increasing numbers $k$ of zero crossings in the bulk (counted after the main peak near the horizon, i.e. not counting zero-crossings from the fractal-like behaviour near the horizon), see Figure~\ref{innerKGmodes}(d). We set $z_{max}=1-z_{min}$ but its precise value is not significant. We have seen already for the horizon modes that there are phase shifts and other changes near the horizon as we change $z_{min}$, but the behaviour is not drastically affected far from the horizon. We see this in the present case also, with the eigenvalues plotted at left against $z_{min}$ for $k=1,2,3,4$ depending only gradually on $z_{min}$ (the plots themselves are interpolated from 10 equally spaced sample points). We also show plots of the $k=4$ modes for the initial, final and middle values of $z_{min}$ to see how their shape changes, with similar comparisons for $k=1,2,3$. There is similar behaviour if we plot the eigenvalues and eigenfunctions against  $z_{min}$ going right down to  the resolution scale $z_{min}={\rm res}$. As a test of robustness, changing to coarser resolution with ten times res but fixed $z_{min}$ does not change the plots.  As for horizon modes, the minimum choice of $z_{min}$ could be seen as a crude model, via the numerical resolution, of possible discretisation due to quantum gravity effects. It should also be possible to see similar quantum gravity effects modelled by noncommutative coordinates.

\section{Concluding remarks}\label{sec:rem}

We have shown that the notion of a classical quantum geodesic flow (i.e. quantum geodesics\cite{Beg:geo,BegMa:geo,BegMa:cur,BegMa:gra,BegMa:qm} but applied to the commutative algebra $C^\infty(M)$)  provides a new tool that could be useful for General Relativity, in which individual geodesics are replaced by the flow of a density. We looked particularly at the main new ingredient $X$, the geodesic velocity field obeying (\ref{velocityflow}), which is more fundamental in that it determines the density  flow via (\ref{densityflow}) rather than the other way around. This was contrary to intuition and required an approach on how to choose $X(0)$ initially. Ultimately, this is a free choice, but we proposed that a natural option is to require ${\rm div}(X(0))=0$ in the interior of a region of interest, coupled with boundary conditions expressing the initial flux of particles in or out of the region.  We then set up a statistical model of a large number of actual geodesics, interpolated as a density, where each particle starts off with velocity $X(0)$ at its location. Their somewhat crudely interpolated tangents then gave a  rough idea of the interpolated $X_{stat}(s)$ broadly comparable in the centre of the region (where the particle density is high enough) with $X(s)$ obtained by solving (\ref{velocityflow}). We could also compare the $\rho(s)$ obtained by solving (\ref{densityflow}) with the interpolated density  $\rho_{stat}(s)$ from the geodesic particle locations when each was evolved by its own proper time $s$. This was good fit but with differences attributed to a change of shape of the peak not picked up in the statistical modelling. This helps to justify $s$ in the geodesic flow theory as some kind of collective proper time. 

After this proof of concept in Section~\ref{sec:stat}, we continued in Section~\ref{sec:examples} to study the geodesic flow theory as a potential new tool for General Relativity in its own right. Of particular interest is the idea that $\rho=|\psi|^2$ for an amplitude $\psi$, and we explored this. We looked at the flow of a bump through the horizon, the flow of a wave-packet $\psi$ and the flow of two colluding $\psi$-bumps of opposite sign (they collide to a dipole). The density profile of the latter is different from the single bump obtained when two $\rho$-bumps collide sufficiently closely. In other words,  the radical hypothesis that (in some circumstances) we might replace a density over spacetime by a wave-function over spacetime could in principle be testable. A significant open problem would be to extend our analysis to include angular variables, since we worked with functions of $U,V$ only (i.e. radial motion) for the numerical integrations to be manageable. Including angular motion would, for example, allow one to demonstrate a gravitational Bohm-Aharonov effect in which a wave function $\psi$ flows around the black-hole and interferes with itself. We also had significant (but manageable) noise problems requiring new techniques (as well as more computing power) to address. These would be interesting directions on the applied side.

In Section~\ref{sec:kg}, we looked at a different theory also involving functions $\phi$ on spacetime, but these are now understood as on or off-shell Klein-Gordon (i.e. scalar) fields. Here the applicable flow was (\ref{KGflow})  and we looked at a particular limit of it called `pseudo-quantum mechanics' since it looks like quantum mechanics but with respect to the collective time $s$. This was already examined in the exterior of a black-hole in \cite{BegMa:qm} but it was a good check of the covariance of the formalism that we obtained the same behaviour now in  Kruskal-Szekeres coordinates. The constant energy modes of interest (with respect to co-ordinate time) now became modes of the form $(-w)^{-{\rm i}p}\psi(z,\theta,\phi)$ where $z=UV$, $w=U/V$ and $p$ is a real parameter, and we again focussed on the case without angles. This led to an elegant Schr\"odinger-like equation (\ref{kgqm}) allowing us to study in more detail the novel process in \cite{BegMa:qm} where a region of density or amplitude outside the horizon moves towards it and ends up producing `horizon modes' bunched at the horizon and oscillating with arbitrarily small wavelength as we approach it.

We further argued that a full understanding that includes the horizon itself (rather than cutting off the domain just above it) would need regularisation such as by discretisation of the spacetime or by making it noncommutative, with the closest cut-off in our case being the resolution of the numerical solver itself. Our initial results from this point of view suggest some kind of `skin' at the horizon as in \cite{Ma:alm} and should be looked at more systematically.  The horizon modes also deserve to be studied further in relation to the flow of information (where it was shown in \cite{BegMa:qm} that the probability density entropy increases throughout the process in which a Gaussian bump ends up turning into these horizon modes). We also found stationary modes (i.e. actual solutions of the Klein-Gordon equations) that resemble atomic states, again as in \cite{BegMa:qm}. This time, using Kruskal-Szekeres coordinates, we were able to find analogous results for both  horizon modes and  atomic modes {\em inside} the black-hole. Atom-like states inside the black-hole could be of interest as internal structure is not visible outside it. As we are looking for stationary modes under the evolution/solving the Klein-Gordon equation, the physical role here could also be more direct. A new feature was that the atomic modes inside the black-hole have a discrete spectrum if we cut off the interior region just above the horizon and impose mixed boundary conditions at the $z=1$ singularity. The eigenvalues in the discrete spectrum depend on the cutoff but with a similar pattern of behaviour right down to the resolution scale. Hence, quantum-gravity effects at the horizon in the form of discretisation or finite resolution directly determine the spectrum of the atomic modes inside the black hole. 

Clearly, our methods could also be applied to other spacetimes. Here, Klein-Gordon flow on FLRW spacetimes was studied in \cite{BegMa:flrw} but the more basic classical geodesic flows which were our main focus in the present work (plus the option of an underlying wave function) should be looked at in such models in relation to cosmology. It would also be important to understand the interaction of our methods with relativistic fluid mechanics (as touched upon in the introduction). Last but not least, how the theory relates to optimal transport and to the (well-known, but different) theory of geodesic flows on the normal bundle of a manifold should be explored further. For a noncommutative algebra such flows have been proposed in \cite{Con95} to be generated by $|D|$ where $D$ is an abstract `Dirac operator' in Connes' notion of a spectral triple. It could therefore be interesting to construct them on the Heisenberg algebra or algebra of differential operators.

\section*{Declarations}

\noindent{\bf Funding:}  KK was supported by DFG project grant 515782239. SM was supported by a Leverhulme Trust project grant RPG-2024-177. 

\medskip
\noindent{\bf Data availability:} Data sharing is not applicable as no data sets were generated or analysed during the current 
study. Software code used for the numerical studies is available online\cite{KM:code}.

\medskip
\noindent{\bf Conflict of Interest:} The authors have no competing  interests to declare that are relevant to the content of this article.

\end{document}